\documentclass[traditabstract]{aa} 
\usepackage{graphicx}
\usepackage{lscape}
\usepackage{natbib}
\usepackage{colortbl}
\usepackage{color}
\usepackage{array}
\usepackage{amsfonts,amssymb,amsmath}
\usepackage{multirow}
\usepackage{afterpage}

\newcommand{\hi}{\mbox{H\,{\sc i}}} 
\newcommand{\hii}{\mbox{H\,{\sc ii}}}

\newcommand{\oi}{\mbox{O\,{\sc i}}} 
\newcommand{\ovi}{\mbox{O\,{\sc vi}}} 
\newcommand{\fei}{\mbox{Fe\,{\sc i}}}
\newcommand{\feii}{\mbox{Fe\,{\sc ii}}}
\newcommand{\feiii}{\mbox{Fe\,{\sc iii}}}
\newcommand{\feiv}{\mbox{Fe\,{\sc iv}}}

\newcommand{\siii}{\mbox{Si\,{\sc ii}}} 
\newcommand{\sii}{\mbox{Si\,{\sc i}}} 
\newcommand{\siiii}{\mbox{Si\,{\sc iii}}} 
\newcommand{\siiv}{\mbox{Si\,{\sc iv}}} 
\newcommand{\alii}{\mbox{Al\,{\sc ii}}} 
\newcommand{\aliii}{\mbox{Al\,{\sc iii}}} 
\newcommand{\nv}{\mbox{N\,{\sc v}}}
\newcommand{\niii}{\mbox{Ni\,{\sc ii}}}
\newcommand{\mgii}{\mbox{Mg\,{\sc ii}}}
\newcommand{\mgi}{\mbox{Mg\,{\sc i}}}
\newcommand{\cii}{\mbox{C\,{\sc ii}}} 
\newcommand{\ciii}{\mbox{C\,{\sc iii}}} 
\newcommand{\crii}{\mbox{Cr\,{\sc ii}}} 
\newcommand{\znii}{\mbox{Zn\,{\sc ii}}} 
\newcommand{\civ}{\mbox{C\,{\sc iv}}} 

\def\grb{GRB\,080310}
\def\lsim{\mathrel{\hbox{\rlap{\lower.55ex \hbox {$\sim$}}\kern-.0em
\raise.4ex \hbox{$<$}}}} 
\def\gsim{\mathrel{\hbox{\rlap{\lower.55ex \hbox {$\sim$}}\kern-.0em
\raise.4ex \hbox{$>$}}}} 
\def\lya{Ly$\alpha$}

\def\ion#1#2{#1$\;${\small\rm\@Roman{#2}}\relax}

\def\kms{km s$^{-1}$}
\def\1star{$^{\star}$}
\def\2star{$^{\star\star}$}
\def\3star{$^{\star\star\star}$}
\def\4star{$^{\star\star\star\star}$}

\begin{document}

\title{Rapid-response mode VLT/UVES spectroscopy \\of super iron-rich gas exposed to GRB\,080310\thanks{\protect Based on Target-Of-Opportunity observations carried out in service mode using the Very Large Telescope Rapid-Response Mode under programme ID 080.D-0526, P.I. Vreeswijk, with the Ultraviolet and Visual Echelle Spectrograph (UVES) installed at the Nasmyth-B focus of the VLT, Unit 2, Kueyen, operated by the European Southern Observatory (ESO) on Cerro Paranal in Chile.}}\subtitle{Evidence of ionization in action and episodic star formation in the host}

\author{A. De Cia 
       \inst{1}
       \and
       C. Ledoux\inst{2}
             \and
       A.J. Fox\inst{2,3}
       \and
       P.M. Vreeswijk\inst{1}
       \and
       A. Smette\inst{2}
              \and
       \\P. Petitjean\inst{4}
        \and
       G. Bj\"{o}rnsson\inst{1}
       \and
      J.P.U. Fynbo\inst{5}
             \and 
      J. Hjorth\inst{5}
      \and
       P. Jakobsson\inst{1}
       }

\institute{
Centre for Astrophysics and Cosmology, Science Institute, University of Iceland, Dunhagi 5, 107 Reykjavik, Iceland\\
\email{annalisa@raunvis.hi.is}
\and
European Southern Observatory, Alonso de C\'ordova 3107, Casilla 19001, Santiago 19, Chile
\and
STScI, 3700 San Martin Drive, Baltimore, MD 21218, USA
\and
Institut d'Astrophysique de Paris, Universit\'e Paris 6, 98bis Boulevard Arago, 75014, Paris, France
\and
Dark Cosmology Centre, Niels Bohr Institute, University of Copenhagen, 2100 Copenhagen, Denmark
}

 \date{Received MM gg, 2012; accepted MM gg, 2012}

\abstract
{We analyse high-resolution near-UV and optical spectra of the afterglow of \grb{}, obtained with the Very Large Telescope Ultraviolet and Visual Echelle Spectrograph (VLT/UVES), to investigate the circumburst environment and the interstellar medium of the gamma-ray burst (GRB) host galaxy. The VLT rapid-response mode (RRM) enabled the observations to start only 13 minutes after the \textit{Swift} trigger and a series of four exposures to be collected before dawn. A low neutral-hydrogen column-density (log\,$N$(\hi{}) $=18.7$) is measured at the host-galaxy redshift of $z=2.42743$. At this redshift, we also detect a large number of resonance ground-state absorption lines (e.g., \cii{}, \mgii{}, \alii{}, \siii{}, \crii{}, \civ{}, \siiv{}), as well as time-varying absorption from the fine-structure levels of \feii{}. Time-varying absorption from a highly excited \feiii{} energy level ($^7$S$_{3}$), giving rise to the so-called UV\,34 line triplet, is also detected, for the first time in a GRB afterglow. The \crii{} ground-state and all observed \feii{} energy levels are found to depopulate with time, whilst the \feiii{} $^7$S$_{3}$ level is increasingly populated. This absorption-line variability is clear evidence of ionization by the GRB, which is for the first time conclusively observed in a GRB afterglow spectrum. We derive ionic column densities at each epoch of observations by fitting absorption lines with a four-component Voigt-profile model. We perform CLOUDY photo-ionization modelling of the expected pre-burst ionic column densities, to estimate that, before the onset of the burst, [C/H] $=-1.3\pm0.2$, [O/H] $<-0.8$, [Si/H] $=-1.2\pm0.2$, [Cr/H] $=+0.7\pm0.2$, and [Fe/H] $=+0.2\pm0.2$ for the integrated line profile, indicating strong overabundances of iron and chromium. For one of the components, we observe even more extreme ratios of [Si/Fe] $\leq-1.47$ and [C/Fe] $\leq-1.74$. These peculiar chemical abundances cannot easily be explained by current models of supernova yields. They are indicative of a low dust content, whilst dust destruction may also contribute to the marked Fe and Cr overabundances. The overall high iron enhancement along the line-of-sight suggests that there has been negligible recent star formation in the host galaxy. Thus, the occurrence of a GRB indicates that there has been episodes of massive star formation in the GRB region.}

\maketitle

\section{Introduction}
\label{sec intro}

Gamma-ray burst (GRB) afterglows, which are produced by the shock between the burst jet and either the surrounding interstellar medium (ISM) or stellar wind, radiate their extremely powerful and featureless continuum through their host galaxies \citep[see][for reviews]{Piran04,Meszaros06}. This offers a unique opportunity to study the properties of such distant and faint galaxies. In particular, optical and near-ultraviolet (near-UV) spectroscopy of long-duration ($t_{\rm obs}>2$~s) GRBs reveals the chemical composition of the gas along the line-of-sight through absorption lines imprinted in the afterglow spectrum \citep[e.g.,][]{Prochaska07}. However, the vast majority of spectra have to date been taken at low resolution \citep[e.g.,][]{Fynbo09,Christensen11}, hampering metallicity studies.

At the high-resolution end ($R\sim45,000$), the Very Large Telescope (VLT) Ultraviolet and Visual Echelle Spectrograph \citep[UVES,][]{Dekker00} and the Keck High-Resolution Echelle Spectrometer \citep[HIRES,][]{Vogt94} allow detailed studies of the metal abundances, physical states, and gas kinematics within the host galaxies and possibly lower-redshift intervening systems \citep{Fiore05,Prochaska07,Vreeswijk07,Fox08,Ledoux09,Vergani09}, albeit for a small sample of bright afterglows. Among the six \textit{Swift}-era GRB afterglows with metallicity determinations from VLT/UVES spectroscopy, four absorbers have low metallicities, [X/H$]<-1$, where X is taken to be either Zn or S. These generally have low dust depletion factors, [X/Fe], indicating a low dust content, while in one case a higher [S/Fe] ratio is suggestive of an $\alpha$-element enhancement \citep{Ledoux09}. On the other hand, \citet{Prochaska07} found that [$\alpha$/Fe] and [Zn/Fe] tend to be higher in GRB absorbers than along quasi-stellar object (QSO) lines-of-sight, based on high- and low-resolution spectroscopy of a sample of 16 absorbers, indicating significant contributions from massive stars and/or differential dust depletion. However, the current GRB absorber samples are still too limited in size and possibly too biased to determine what the typical abundances of GRB host galaxies are.

A larger collection of metal column densities in 29 GRB absorbers, drawn from both high- and low-resolution spectroscopy \citep{Schady11}, reveals that the ranges of relative chemical abundances measured in GRB absorbers are generally similar to those determined along QSO lines-of-sight. In this paper, we present peculiar metal abundances in the host galaxy of \grb{} and discuss their possible origin.

Thanks to time-resolved high-resolution spectroscopy, the variability of fine-structure lines has been detected in GRB absorbers and explained by photo-excitation induced by the incident GRB afterglow flux \citep[i.e., UV pumping,][]{Vreeswijk07,Prochaska06a}. On the other hand, the variability of resonance metal lines (i.e., those associated with the ground-state level) is expected in the case of photo-ionization \citep[e.g.,][]{Perna98} but has never been conclusively detected until now \citep[the possible variability of \lya{} towards GRB\,090426 was reported by][]{Thone11}. Here we present the detection of significant variability in both fine-structure and resonance metal absorption lines in the afterglow spectrum of \grb{}. In addition, a triplet of lines arising from a highly excited level of \feiii{} is observed in this absorber, which is a primer in a GRB afterglow spectrum. As we discuss in this paper, these observations clearly show that photo-ionization by the GRB has taken place. Self-consistent photo-excitation and ionization modelling to determine the GRB-absorbing cloud distance will be presented in a companion paper \citep[][hereafter referred to as paper II]{Vreeswijk12}.

This paper is organized as follows. In Sect.~2, we report on the observations and the data reduction. In Sect.~3, we derive ionic column densities from Voigt-profile fitting of identified absorption lines and in Sect.~4 we present CLOUDY photo-ionization modelling. We discuss our results in Sect.~5 and conclude in Sect.~6. Throughout the paper, we adopt ions cm$^{-2}$ as the linear unit of column density $N$. The relative abundance of two chemical elements, $X$ and $Y$, is defined as $\left[X/Y\right] \equiv \log{\frac{N(X)}{N(Y)}} - \log{\frac{N(X)_\odot}{N(Y)_\odot}}$. We estimate its uncertainty by adding in quadrature the errors in the observed column densities and in the solar abundances involved. For the reference solar abundances appearing in the second term of this formula, we follow the recommendations of \citet{Lodders09}, adopting either their meteoritic estimates, the photospheric values of \citet{Asplund09}, or the average of these both. The solar abundances used in this paper are summarized in Table~\ref{tab solar}.

\begin{table}
\caption{Solar abundances used in this paper}
\centering
\begin{tabular}{ @{}c c c | c c c @{}}
\hline \hline
\rule[-0.2cm]{0mm}{0.8cm}

Element  & $A(\rm El)\pm\sigma_A$ $^a$ & S$^b$ & Element  & $A(\rm El)\pm\sigma_A$ $^a$ & S$^b$\\
\hline
\rule[-0.0cm]{0mm}{0.4cm}

H & $\equiv12$       &  s   & Cr  &  $5.64\pm0.04$ &  a \\
C &  $8.43\pm0.05$ &  s &  Fe  & $7.47\pm0.04$ & a\\
O & $8.69\pm0.05$  &  s & Ni  & $6.21\pm0.04$ & a\\
\rule[-0.2cm]{0mm}{0.4cm}
Si  & $7.51\pm0.01$ & m & Zn & $4.63\pm0.04$  &  m\\


\hline\hline
\end{tabular}
 \tablefoot{$^a$ Abundances $A(\rm El) \equiv \log N(\rm El)/N(\rm H) + 12$ are taken from \citet{Asplund09} following the recommendations of \citet{Lodders09}. $^b$ Source of the estimate: solar photosphere (s), meteorites (m), or the average between the two (a).}
\label{tab solar}
\end{table}

\begin{table*}
\caption{Log of VLT/UVES observations}
\centering
\begin{tabular}{ @{}l | c c @{\hspace{3mm}} c @{\hspace{3mm}} c @{\hspace{3mm}} c c @{\hspace{2mm}} c @{\hspace{2mm}} c @{\hspace{3mm}} c c @{}}
\hline \hline
\rule[-0.2cm]{0mm}{0.6cm}
Ep. & $t_{\rm start}^a$     & $\Delta\,t^b$& $t_{\rm exp}$ &  Dic.   &  Setting   & Coverage & FWHM$^c$   & Mean & S/N$^d$ & $R=\frac{\lambda}{\Delta\lambda}$\\
\rule[-0.2cm]{0mm}{0.6cm}
& UT  &   (min)    &   (min)  &    \#   & $\lambda_{\rm c}$ (nm)&$\lambda_{\rm obs}$ (nm)& blue/red ($\arcsec$) &airmass & blue/red &  blue/red\\  
\hline

 & & & & & & & & & & \\
I& 08:51:05   &   14.55    &     3   &      1   & 346+580 & 303--388; 476-- 684  & 1.1/1.0 & 1.12 &  1.9--2.0/4.0--4.5 & 48,500/45,600\\
II& 08:56:46   &   21.23    &     5   &      2   & 437+860  & 373--499; 660--1060 & 1.2/0.9 & 1.12 & 3.4--4.7/2.7--8.4 & 47,900/45,300\\
III& 09:04:20   &   31.25    &    10   &      1   & 346+580 & 303--388; 476-- 684  & 1.5/1.3 & 1.14 & 3.1--3.7/6.8--7.6 & 48,500/45,600\\
IV& 09:16:36   &   50.06    &    23   &      2   & 437+860 & 373--499; 660--1060 & 1.6/1.2 & 1.17 & 6.0--8.7/4.8-14.3 & 47,900/45,300\\
 & & & & & & & & & & \\

\hline \hline
\end{tabular}
\tablefoot{$^a$ March 10 2008. $^b$ Mid-exposure time after the burst event in the observer's frame. $^c$ Full width at half maximum of the spatial profile in the two-dimensional spectrum. $^d$ Signal-to-noise per pixel; min--max ranges over the spectral regions where absorption lines are observed.}
\label{tab_log}
\end{table*}

\section{Observations and data reduction}

GRB\,080310 triggered the \textit{Swift} Burst Alert Telescope (BAT) on March 10 2008 at UT 08:37:58 \citep{Cummings08}. One and a half minutes later, the \textit{Swift} X-Ray Telescope (XRT) provided an X-ray afterglow position of RA $=14^\mathrm{h}40^\mathrm{m}13.5^\mathrm{s}$ and Dec $=-00^\circ10\arcmin32.1\arcsec$ (J\,2000), with an uncertainty of $5\arcsec$. This detection led us to trigger the RRM of VLT/UVES, as set up by our team, to slew the Kueyen telescope to the target automatically. Once the object was centred on the slit, the UVES observations started (UT 08:51:01), i.e., 13 minutes and three seconds after the BAT trigger \citep{Vreeswijk08}.

Four exposures were taken, of increasing durations to compensate for the fading afterglow brightness, alternating between dichroic \#1 and dichroic \#2 with standard UVES settings $346+580$ and $437+860$, respectively. In this way, the near-UV/optical range was basically entirely covered via two consecutive exposures. The instrument settings used have three overlapping regions of widths $\sim150$ \AA{}, $\sim230$ \AA{}, and $\sim240$ \AA{}, which were observed at all four epochs, and two $\sim100$ \AA\ gaps in the red owing to the physical separation between the two red CCDs. The CCDs were set to bin pixels $2\times2$ and the spectrograph slits to be $1\arcsec$ wide, providing a mean resolving power $R=\lambda/\Delta\lambda\sim46,400$ (full width at half maximum FWHM$\sim6.5$ \kms{}).

The data were reduced using the ESO UVES pipeline v2.9.7 based on MIDAS \citep{Ballester00} and the wavelength scale was then transformed to the heliocentric rest-frame. As a consistency check on the quality of the reduced spectra, an independent data reduction was also performed using the UVES pipeline v3.4.5 based on the Common Pipeline Library (CPL).\footnote{ESO MIDAS and CPL-based UVES pipelines are available at http://www.eso.org/sci/data-processing/software/pipelines/}, providing very similar quality products. The observing log and details on the observations are presented in Table~\ref{tab_log}.

\begin{table}
\centering
\caption{Transitions simultaneously modelled with a Voigt profile.}
\begin{tabular}{ c | l l r }
\hline \hline
\rule[-0.2cm]{0mm}{0.8cm}
Epochs & Ion  & Level & Transitions $\lambda$ (\AA{}) \\
\hline
\rule[-0.0cm]{0mm}{0.4cm}
 & \siii{} &  $\,^2$P$_{1/2}$ &  1526 \\
 & \feii{} &  $\,^6$D$_{9/2}$ &  1608\\
I, III  & \feiii{} &  $\,^5$D$_{4}$   &  1122 \\
\rule[-0.2cm]{0mm}{0.4cm}
 & \feiii{}$^{17*\,a}$ & $\,^7$S$_{3}$    &  UV\,34: 1895, 1914, 1926 \\
\hline
\rule[-0.0cm]{0mm}{0.4cm}
 & \cii{}  &  $\,^2$P$_{1/2}$&  1334 \\
 & \siii{} &  $\,^2$P$_{1/2}$ &  1190, 1193, 1260, 1304 \\
 & \siii{}* & $\,^2$P$_{3/2}$    &  1264, 1265 \\
 & \feii{} & $\,^6$D$_{9/2}$  &   2344, 2374, 2382$^b$, 2586   \\
II, IV  & \feii{}* & $\,^6$D$_{7/2}$    &  2333, 2365, 2389, 2396, 2612 \\
 & \feii{}** & $\,^6$D$_{5/2}$     & 2349, 2399, 2405 \\
 & \feii{}*** & $\,^6$D$_{3/2}$     &  2338, 2407, 2411 \\
& \feii{}****&  $\,^6$D$_{1/2}$  & 2411, 2414 \\
\rule[-0.2cm]{0mm}{0.4cm}
 & \feiii{}&  $\,^5$D$_{4}$         &   1122  \\
\hline \hline
\end{tabular}
\tablefoot{$^a$ Seventeenth energy level above the ground-state. $^b$ Only the redder component is fitted since the absorption profile is saturated in bluer regions.}
\label{tab lines to fit}
\end{table}

\section{Absorption-line analysis}

\begin{figure*}
   \centering
   \includegraphics[width=180mm,angle=0]{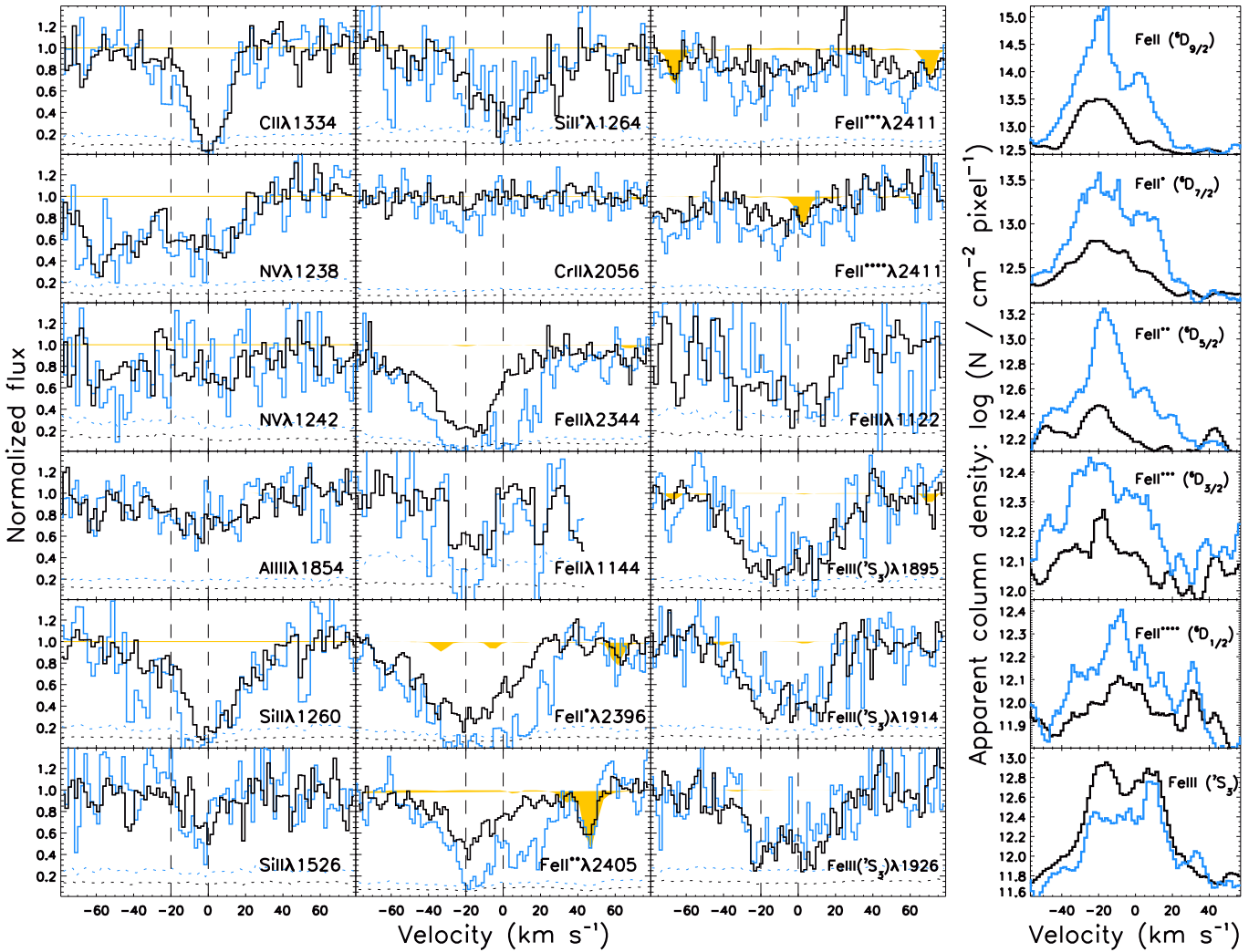}
      \caption{\textbf{Left:} a selection of line profiles observed at different epochs (epochs I and II in blue; epochs III and IV in black). Telluric features are highlighted in yellow. The dashed lines mark the position of the strongest component of all the \feii{} lines and the strongest component of \siii{}, \cii{} and \aliii{} lines, respectively. \textbf{Right:} the apparent column-density profiles (smoothed with a boxcar of five pixels) of the \feii{} and \feiii{} levels, obtained by combining together different transitions that probe a particular level, show a clear evolution with time. In particular, while all \feii{} levels depopulate, \feiii{} shows the opposite trend.}         \label{fig: profiles var}
   \end{figure*}

\subsection{Detected lines}

We first combined the spectra obtained at the four different epochs of observations, to achieve the highest signal-to-noise ratio (S/N) for the identification of absorption features. Starting with the detections of the strongest doublet lines and confirming these with associated resonance lines detected at the same redshift, we identified absorption lines belonging to eight distinct systems. We associated the highest redshift system, at $z=2.42743$, with the GRB host galaxy. The presence of \feii{} fine-structure lines in this system confirms this association, since their origin is most likely linked to the GRB \citep[e.g.,][]{Vreeswijk07}. In this paper, we focus our analysis on the absorption within the host galaxy, and exclude the absorption lines associated with the seven intervening systems. The latter are listed in Table~\ref{tab inter} (see Appendix) with equivalent width ($W$) estimates.

At the GRB redshift, we confirm the presence of absorption features such as \lya{}, and highly ionized species, such as \civ{}, \siiv{}, \nv{}, and \ovi{}, as reported in \cite{Fox08} and \cite{Ledoux09}. We also detect low-ionization absorption lines, such as \cii{} $\lambda$ 1334, \alii{} $\lambda$ 1670, \aliii{} $\lambda\lambda$ 1854, 1862, \mgii{} $\lambda\lambda$ 2796, 2803, and a blended \oi{} $\lambda$ 1302, in addition to \siii{}, \feii{} ground-state and fine-structure, and \feiii{} ground-state transitions. Moreover, three lines displaying similar profiles are clearly detected, but cannot be associated with any transition reported for GRB lines-of-sight to date, nor any of the intervening systems. We identify these transitions as the \feiii{} UV\,34 triplet $\lambda\lambda\lambda$ 1895, 1914, 1926. We discuss these lines and their excited-level origin in detail in Sect.~\ref{discussion}. Remarkably, the line profiles at different epochs reveal the clear variability in both the \feii{} and \feiii{} transitions. Fig.~\ref{fig: profiles var} shows the (non-) variability in some representative transitions, in the left panel, and the \feii{} and \feiii{} apparent column densities (derived by combining together the transitions of each level), in the right panel. A decrease in the \feii{} apparent column density is evident in all the levels, while the excited \feiii{} $^7$S$_3$ population increases with time. We discuss the evidence for the variability of \feii{} and \feiii{} and other species further in Sect.~\ref{sec evidence var}.

\subsection{Voigt-profile fitting}
\label{spec var}

Table~\ref{tab lines to fit} lists all metal lines, including both ground-state and excited-level transitions, that are clearly detected yet not strongly saturated and therefore suitable for the Voigt-profile fit. A first look at the absorption lines identified in the spectrum reveals that their profiles are not identical for all the ions. In particular, the strongest component of all the \feii{} lines appears broader and blue-shifted by $\sim20$ \kms{} with respect to the strongest component of the line profile of the other low-ionization species, such as \siii{}, \cii{}, and \aliii{} (see, e.g., Fig.~\ref{fig: profiles var}). 

Before investigating the physical origin of this shift, we need to ensure it is not an artefact caused by the wavelength calibration. In this check, a first point of reference is the \feiii{} $\lambda$ 1122 transition, which is observed with a similar profile at the same wavelength, despite the noise, in all four spectra in the overlapping region between the two different settings. Second, while most of the \feii{} transitions are observed in the red arm spectra obtained with dichroic \#2, the bluer \feii{} $\lambda$ 1608 shows the same \feii{} profile in the spectra taken with dichroic \#1. Moreover, the \feii{} $\lambda$ 1608 and the \siii{} $\lambda$ 1526 lines are observed simultaneously within the same spectral region, indicating that the wavelength difference between the strongest iron and silicon components is real. In addition, the \feii{} $\lambda\lambda$ 1144, 1096 transitions that are observed in the blue arm spectra obtained with the dichroic \#1 and \#2 show the same kinematics as the red arm \feii{} (stronger component at $-20$ \kms{}). In particular, the \feii{} $\lambda$ 1144 shows clear variability (see Fig.~\ref{fig: profiles var}), confirming the line identification. Finally, a synthetic telluric spectrum (see next Section) closely matches the observations across the full spectral range, proving the reliability of our wavelength calibration.

The absorption-line profiles of these low-ionization species differ significantly from the high-ionization ones analysed by \citet{Fox08}, with \civ{}, \ovi{}, and \siiv{} having much broader line profiles with blue-shifted wings (up to 220 \kms{} for \civ{}). Thus, we can exclude a direct association between the high-ionization wings and the low-ionization components. A physical separation between the low- and high-ionization species, and therefore different line profiles, is not unexpected, as they may trace different regions of the ISM owing to their different ionization potentials.

\subsubsection{\cii{}, \aliii{}, \siii{}, \feii{}, and \feiii{}}

To derive the column densities of the metals and their evolution with time, we modelled the observed lines with the Voigt-profile fitting software VPFIT\footnote{\texttt{http://www.ast.cam.ac.uk/$\sim$rfc/vpfit.html}}. We first normalized the spectrum locally using carefully selected featureless regions, around each line, by fitting a zero- or first-order polynomial to these regions. We considered the best-fit continuum level plus and minus 0.5 times the noise root mean square in the adjacent continuum (upper and lower continua), following \citet{Ledoux09}. This allowed us to take into account the normalization error in the estimate of the error in the column densities. The normalization error tends to dominate over the formal Voigt-profile fit error for weak transitions.

To avoid false identifications caused by sky absorption lines, we created synthetic telluric absorption-line spectra (see, e.g., Fig.~\ref{fig: profiles var}). These telluric spectra were obtained with an IDL routine acting as a wrapper to the Reference Forward Model (RFM) version 4.28, a line-by-line radiative transfer code\footnote{\texttt{http://www.atm.ox.ac.uk/RFM/}}. The HITRAN'2008 \citep{Rothman09} database provided the molecular absorption-line parameters. The atmospheric profiles describing the pressure, temperature, and humidity at 20 different atmospheric layers were obtained from the Air Resources Laboratory READY Archived Meteorology website\footnote{\texttt{http://ready.arl.noaa.gov/READYinfo.php}}. Following \citet{Smette10}, we calculated synthetic telluric absorption-line spectra for different amounts of precipitable water vapour. Fitting the synthetic telluric spectra to the observed spectra between 7100 \AA\ and 7450 \AA\ provides a best-fit precipitable water-vapour column of $3.84\pm0.05$ mm. In the Voigt-profile fits, we systematically discarded the spectral regions where the telluric absorption features drop below 99\% of the continuum level.

We treated the spectra taken at each epoch separately, to allow for the detection of absorption-line variability. Most species are observed only at two epochs, since the wavelength coverage differs between epochs. Exceptions are the \feiii{} $\lambda$ 1122 resonance line and its fine structure \feiii{}* $\lambda$ 1124\footnote{We use the non-detection of \feiii{}* $\lambda$ 1124 to put an upper limit on the \feiii{}* population.}, which are covered at all four epochs. To simultaneously fit all the line profiles listed in Table~\ref{tab lines to fit} with the same components, we first focused on the highest S/N (epoch IV) spectrum. We then included the epoch-III spectrum to also cover the \feiii{} UV\,34 triplet.

Noticeably, the lines associated with different ions show different profiles, making their decomposition non-trivial. As a first approach, only the stronger \cii{}, \siii{}, \siii{}*, and \feii{} ground-state lines were considered for the identification of the four velocity components ``a'', ``b'', ``c'', and ``d'', listed in Table~\ref{tab cd}, that are necessary to describe the line profile adequately. Below, we briefly discuss the justification for this decomposition.

\begin{figure*}
   \centering
  	\includegraphics{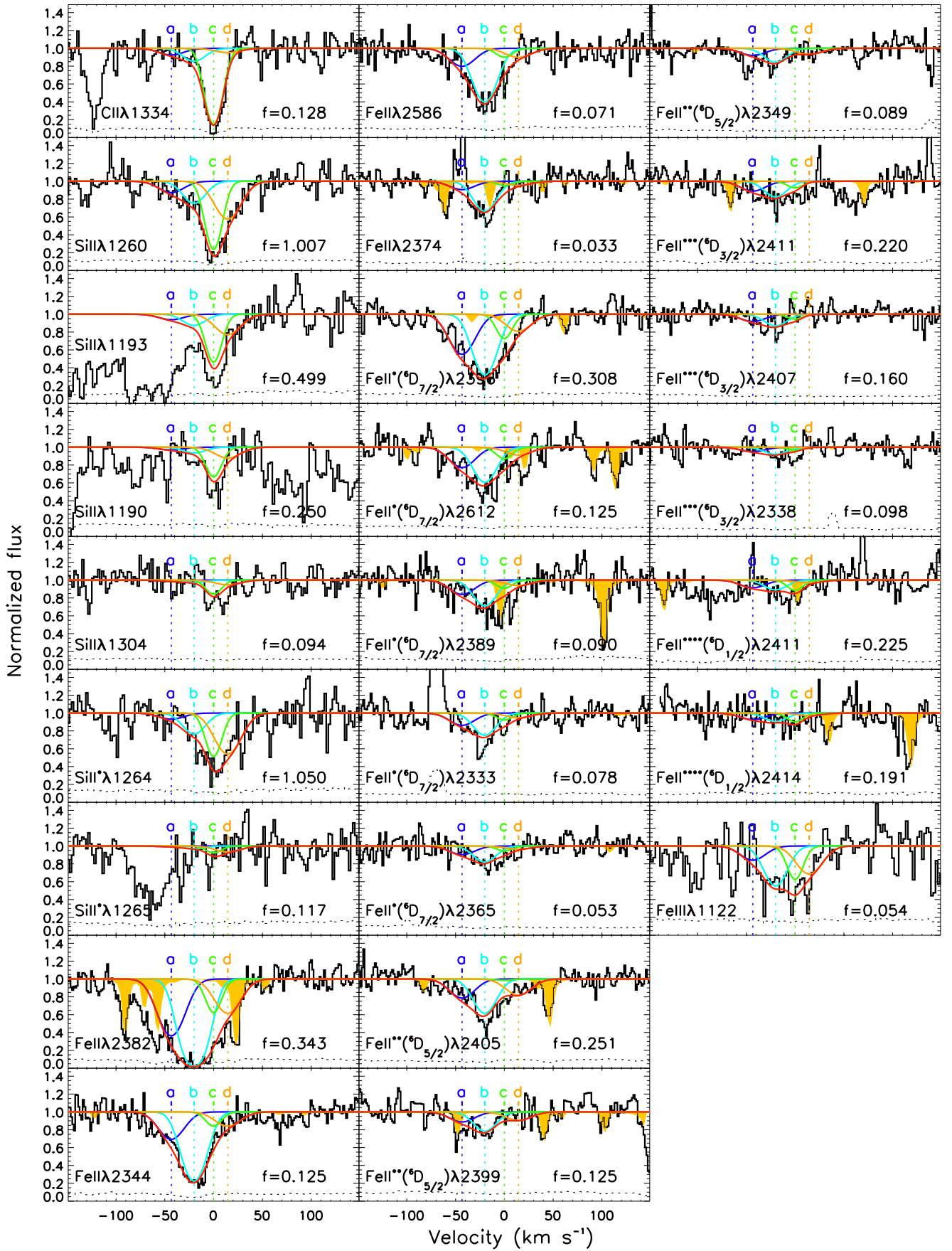}
     \caption{Voigt-profile fit of absorption lines observed at epoch IV. The black solid lines show the observed spectrum, with the associated error indicated by the dashed line at the bottom. The red lines show the combined model fit. Other coloured lines show the individual components. The four components, ``a'', ``b'', ``c'', and ``d'', are labelled accordingly. The transitions are listed in order of decreasing oscillator strength, for each population. Telluric features are highlighted in yellow.}
         \label{fig: fit 4}
   \end{figure*}
 \begin{figure*}
   \centering
  	\includegraphics{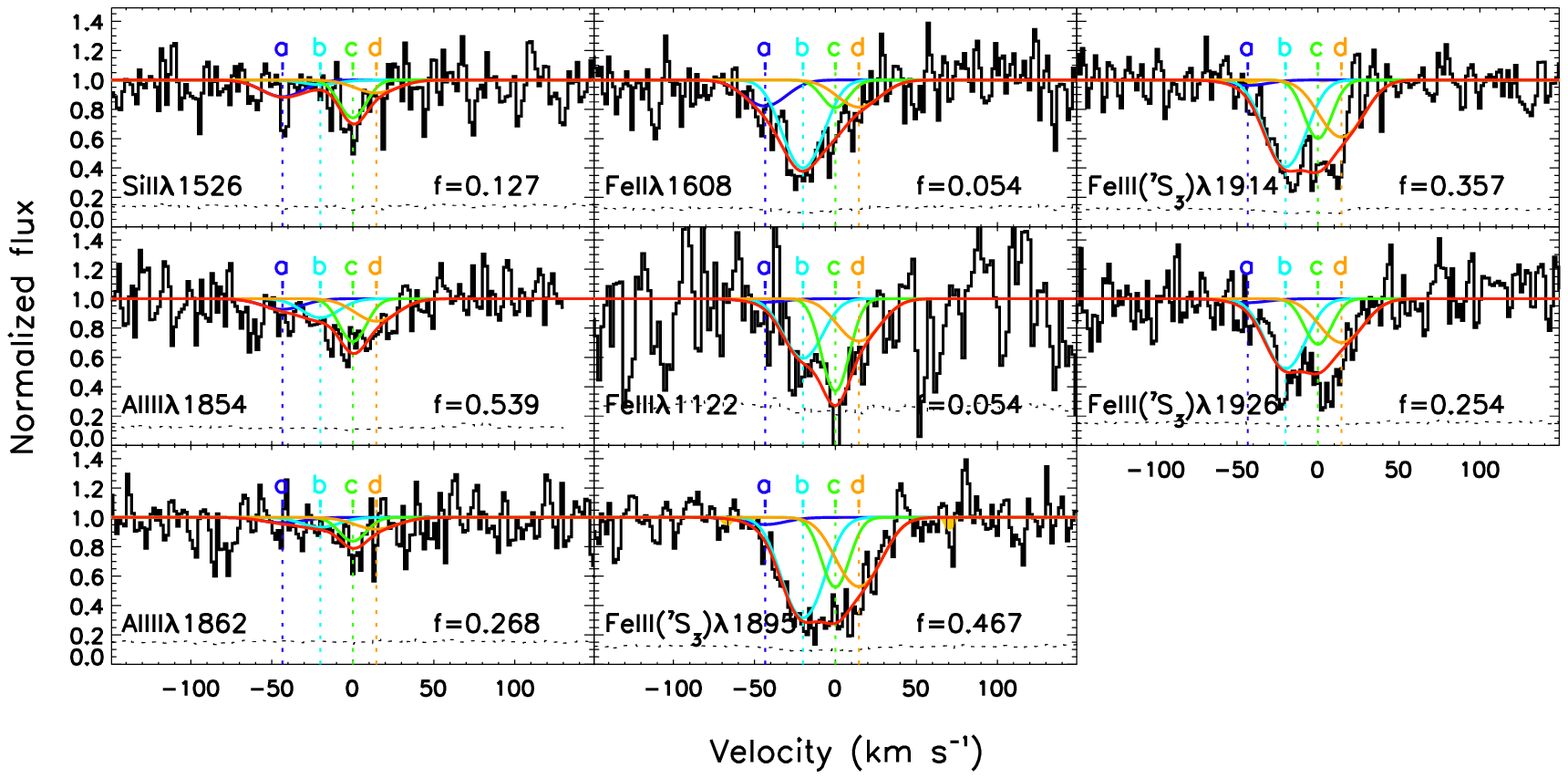}
     \caption{Voigt-profile fit of absorption lines observed at epoch III. The fit decomposition was derived from the highest S/N spectrum (obtained at epoch IV).}
         \label{fig: fit 3}
   \end{figure*}

First, the \cii{} profile is mainly described by a strong narrow ($b=7$ \kms{}) component, which we refer to as component ``c'', at redshift $z=2.42743\pm0.00002$ \footnote{The dominating systematic error, arising from the heliocentric velocity and wavelength calibration, is estimated by i) comparing our spectra with a synthetic telluric one and ii) modelling simultaneously several transitions from different regions of the spectrum, observed with different dichroics and at different epochs.}. We adopt this redshift as our zero-velocity reference point. At this same position, we detect the strongest component of the \siii{} lines, as well as all the other identified transitions. However, despite the strength of component ``c'' in terms of silicon and carbon, iron does not show its stronger absorption at this wavelength position. Second, the \feii{} main contribution, which we refer to as component ``b'', is blue-shifted with respect to the main Si and C components, but is also clearly observed in C and Si. Third, the \feii{} profiles display an even bluer, quite-weak-but-evident, absorption-component ``a'', that is almost undetected or very weak for the other ions. Finally, a redder component, ``d'', widens the profile of the Si lines, although this is less obvious for the \feii{} lines. To investigate the contribution of component ``d'' to \feii{} we include in the fit only the red wing of \feii{} $\lambda$2382, which is otherwise saturated (\feii{} $\lambda$2600 is also saturated). Given the resolution of the instrument of $\sim7$ \kms{}, the individual components derived with the Voigt-profile fit are blended, and therefore the decomposition is not unique. However, the remarkable difference between the \feii{} transitions and the \siii{} and \cii{} lines enables component ``b'' and component ``c'' to be distinguished and therefore more reliably constrained.

We then fixed the positions (redshifts) and widths (velocity-broadening parameter) of components ``a'',``b'', ``c'', and ``d'', reported in Table~\ref{tab cd}, and included all the transitions listed in Table~\ref{tab lines to fit} at their highest S/N (all transitions detected at epoch IV and the \feiii{} transitions at epoch III), in a simultaneous four-component fit of all lines. A pure turbulent broadening (i.e., neglecting thermal broadening) provided the best fit to the line profiles. Figs.~\ref{fig: fit 4} and \ref{fig: fit 3} show the model fit, along with its individual components, convolved with the instrumental resolution; the reduced chi-square is $\chi^2_\nu=1.4$. As mentioned above, the spectral regions affected by telluric features were excluded from the fit.

To combine the errors derived from the Voigt-profile fit with the errors introduced by the normalization, we measured the column densities for the three continuum levels (best fit, lower, and upper) and we add in quadrature the difference between the column densities of the various continua with the formal VPFIT errors. The four-component decomposition is significant for all the transitions in the upper continuum level, i.e. all components are required by the software to describe the line profiles. In the case of a non-detected component in the best-fit continuum, and for weak components with upper errors of more than twice the inferred column density, we used an upper limit corresponding to the largest value between the column density estimate from the upper continuum level and the $3\sigma$ detection limit. For components that are well-constrained in the best-fit continuum, but not detected in the lower continuum, we adopted the expectation value of half the $3\sigma$ detection limit to calculate the lower error. The detection limits for each component were calculated by measuring the uncertainty in the $W$ from the error spectrum\footnote{$\Delta W_{\rm rest}=pixel\,size\,\mbox{[\AA{}]} \times \sqrt{\Sigma_i\,\,normalized\,error_i^2}\,/(1+z)$, where $i$ spans the chosen aperture.} at the position of the stronger transition of each ion, and converting it to column densities assuming the optically thin limit. An aperture of size twice the FWHM was chosen to enclose $\sim98$\% of the surface, for each component found in the Voigt-profile fit. We used FWHM$_i=2\sqrt{2\,\ln2}\,\sigma_i$, where $\sigma_i^2=b_i^2/2$ is the Gaussian variance and $b_i$ is the Doppler-broadening parameter of component $i$. Since the proximity of components ``c'' and ``d'' may cause some degeneracy in the measurement of the individual column densities, we also considered the results of the combined components ``c+d'' in our abundance analysis. The $3\sigma$ detection limit to components ``c+d'' and the total line profile are measured assuming the overall apertures 
\begin{table*}
\caption{Column densities}
\begin{tabular}{ c | l | r r r r r r}
\hline \hline
\rule[-0.3cm]{0mm}{0.8cm}
 &     \multicolumn{1}{r|}{Component}       &    a   &   b   &  c   &  d   &  c+d &  total \\
 \hline
\rule[-0.0cm]{0mm}{0.4cm}
      & \multicolumn{1}{r|}{$z$} & $2.42693$&$2.42720$&$2.42743$&$2.42759$& & \\ 
  &  \multicolumn{1}{r|}{$v$ (\kms{})}& $-43$  & $-20$ & 0$^a$  & 14  \\
  \rule[-0.2cm]{0mm}{0.4cm}
& \multicolumn{1}{r|}{$b^b$ (\kms{})} & $16$  & 13 & 7  & 16  \\
\hline
\rule[-0.3cm]{0mm}{0.8cm}
Epoch & Ion (level) & \multicolumn{6}{c}{log\,$N\pm\,\sigma_{\textrm{log}N}^c$}  \\

 \hline

 \rule[-0.0cm]{0mm}{0.4cm}
  & \aliii{}  &  $<12.76$   &  $<12.62$  &  $<12.21$  &  $<12.74$  &  $<12.85$  &  $12.92^{+0.23}_{-0.15}$  \\
   & \sii{} $[\lambda1562]^d$ &  $<12.68$ &  $<12.64$ &  $<12.53$ &  $<12.68$ &  $<12.68$ &  $<12.99$\\
 & \siii{}  &  $<13.18$   &  $<13.15$  &  $13.24^{+0.13}_{-0.20}$  &  $<13.18$  &  $13.51^{+0.08}_{-0.10}$  &  $13.60^{+0.13}_{-0.11}$  \\
 I & \feii{} ($\,^6$D$_{9/2}$)  &  $<13.58$   &  $14.15\pm0.07$  &  $13.63\pm0.14$  &  $13.74^{+0.17}_{-0.23}$  &  $13.99^{+0.12}_{-0.13}$  &  $14.42\pm0.06$  \\
  & \feiii{} ($\,^5$D$_{4}$)  &  $<14.15$   &  $<13.98$  &  $<13.85$  &  $<14.08$  &  $<14.28$  &  $14.35^{+0.19}_{-0.13}$  \\
 & \feiii{}*$^e$ ($\,^5$D$_{3}$) $[\lambda1124]$&  $<14.16$   &  $<14.12$  &  $<13.98$  &  $<14.17$  &  $<14.17$  &  $<14.44$  \\
\rule[-0.2cm]{0mm}{0.4cm} 
& \feiii{}$^{17}$* ($\,^7$S$_{3}$)  &  $<12.38$   &  $12.87^{+0.08}_{-0.12}$  &  $12.62\pm0.10$  &  $12.88^{+0.11}_{-0.14}$  &  $13.07^{+0.08}_{-0.09}$  &  $13.31^{+0.06}_{-0.07}$  \\

  \hline
 \rule[-0.0cm]{0mm}{0.4cm}
  & \cii{}  &  $<13.06$   &  $13.47^{+0.12}_{-0.18}$  &  $\geq^f14.08$  &  $<13.24$  &  $\geq14.14$  &  $\geq14.22$  \\
 & \cii{}* $[\lambda1335]$ &  $<13.20$   &  $<13.16$  &  $<13.03$  &  $<13.21$  &  $<13.20$  &  $<13.51$  \\
 & \oi{}  &  $<13.75$   &  $<14.40^g$  &  $<13.59$  &  $<13.75$  &  $<13.98$  &  $<14.60$  \\
 & \oi{}* $[\lambda1304]$&  $<13.68$   &  $<13.64$  &  $<13.51$  &  $<13.68$  &  $<13.68$ &  $ <13.98$  \\
  & \siii{}  &  $12.62^{+0.17}_{-0.20}$   &  $12.73^{+0.12}_{-0.13}$  &  $\geq^f13.23$  &  $12.50^{+0.24}_{-0.58}$  &  $\geq13.30$  &  $\geq13.47$  \\
 & \siii{}*  &  $<12.41$   &  $12.43^{+0.16}_{-0.40}$  &  $12.58\pm0.12$  &  $12.53^{+0.18}_{-0.47}$  &  $12.86^{+0.11}_{-0.18}$  &  $13.07^{+0.09}_{-0.13}$  \\
 & \crii{}  &  $<12.84$   &  $13.07^{+0.11}_{-0.18}$  &  $<12.70$  &  $<12.85$  &  $<13.08$  &  $13.33^{+0.12}_{-0.11}$  \\
II & \feii{} ($\,^6$D$_{9/2}$)  &  $13.40^{+0.08}_{-0.09}$   &  $14.10\pm0.04$  &  $13.51\pm0.06$  &  $13.51^{+0.08}_{-0.10}$  &  $13.81^{+0.05}_{-0.06}$  &  $14.34\pm0.03$  \\
 & \feii{}* ($\,^6$D$_{7/2}$)  &  $12.99^{+0.16}_{-0.14}$   &  $13.63^{+0.05}_{-0.06}$  &  $13.01\pm0.10$  &  $13.24^{+0.10}_{-0.08}$  &  $13.44^{+0.07}_{-0.06}$ &  $ 13.91\pm0.04$  \\
 & \feii{}** ($\,^6$D$_{5/2}$)   &  $12.51^{+0.23}_{-0.45}$   &  $13.43\pm0.04$  &  $12.77^{+0.09}_{-0.10}$  &  $13.02^{+0.09}_{-0.11}$  &  $13.22^{+0.07}_{-0.08}$  &  $13.67\pm0.04$  \\
 & \feii{}*** ($\,^6$D$_{3/2}$)  &  $12.73^{+0.15}_{-0.52}$   &  $13.02^{+0.07}_{-0.08}$  &  $12.53^{+0.11}_{-0.13}$  &  $12.54$  &  $12.84\pm0.06$  &  $13.33^{+0.06}_{-0.10}$  \\
 & \feii{}**** ($\,^6$D$_{1/2}$)   &  $<12.60$   &  $12.70^{+0.13}_{-0.18}$  &  $12.49^{+0.13}_{-0.14}$  &  $12.67^{+0.19}_{-0.50}$  &  $12.89^{+0.13}_{-0.24}$  &  $13.18^{+0.09}_{-0.12}$  \\
 & \feii{}***** ($\,^4$F$_{9/2}$) $[\lambda2348]$ &  $<13.19$   &  $<13.15$  &  $<13.03$  &  $<13.19$  &  $<13.21$  &  $<13.56$  \\
 & \feiii{} ($\,^5$D$_{4}$)  &  $<13.79$   &  $<13.76$  &  $<13.62$  &  $14.13^{+0.18}_{-0.24}$  &  $14.24^{+0.15}_{-0.17}$  &  $14.32^{+0.19}_{-0.16}$  \\
 & \feiii{}* ($\,^5$D$_{3}$) $[\lambda1124]$ &  $<13.94$   &  $<13.89$  &  $<13.79$  &  $<13.94$  &  $<13.93$  &  $<14.21$  \\
  & \niii{} $[\lambda1370]$ &  $<13.55$   &  $<13.53$  &  $<13.40$  &  $<13.56$  &  $<13.56$  &  $<13.86$  \\
\rule[-0.2cm]{0mm}{0.4cm}
 & \znii{} $[\lambda2026]$ &  $<12.09$   &  $<12.06$  &  $<11.93$  &  $<12.09$  &  $<12.09$  &  $<12.41$  \\
 \hline

 \rule[-0.0cm]{0mm}{0.4cm}
  & \aliii{}  &  $<12.40$   &  $<12.42$  &  $12.45\pm0.18$  &  $<12.60$  &  $12.83^{+0.08}_{-0.07}$  &  $12.86^{+0.19}_{-0.12}$  \\
   & \sii{} $[\lambda1562]$ &  $<12.42$ &  $<12.38$ &  $<12.25$ &  $<12.42$ &  $<12.42$ &  $<12.73$\\
 & \siii{}  &  $<13.06$   &  $<12.88$  &  $13.09^{+0.10}_{-0.11}$  &  $<13.02$  &  $13.36^{+0.06}_{-0.05}$  &  $13.44^{+0.12}_{-0.08}$  \\
 III & \feii{} ($\,^6$D$_{9/2}$)  &  $13.43^{+0.17}_{-0.22}$   &  $14.06^{+0.04}_{-0.05}$  &  $13.28^{+0.15}_{-0.16}$  &  $13.46^{+0.18}_{-0.23}$  &  $13.68^{+0.13}_{-0.14}$  &  $14.28\pm0.05$  \\
 & \feiii{} ($\,^5$D$_{4}$)  &  $<13.64$   &  $13.96^{+0.14}_{-0.21}$  &  $14.14\pm0.15$  &  $13.83^{+0.25}_{-0.49}$  &  $14.31^{+0.14}_{-0.16}$  &  $14.50\pm0.11$  \\
 & \feiii{}* ($\,^5$D$_{3}$) $[\lambda1124]$ &  $<13.80$   &  $<13.77$  &  $<13.64$  &  $<13.81$  &  $<13.81$  &  $<14.09$  \\
\rule[-0.2cm]{0mm}{0.4cm}
 & \feiii{}$^{17}$* ($\,^7$S$_{3}$)  &  $<12.27$   &  $13.16^{+0.03}_{-0.04}$  &  $12.79\pm0.05$  &  $12.95^{+0.07}_{-0.09}$  &  $13.18\pm0.05$  &  $13.48\pm0.03$  \\

 \hline
 \rule[-0.0cm]{0mm}{0.4cm}
 & \cii{}  &  $<12.96$   &  $12.99^{+0.20}_{-0.56}$  &  $\geq^f14.08$  &  $<12.85$  &  $\geq14.10$  &  $\geq14.14$  \\
 & \cii{}* $[\lambda1335]$ &  $<12.89$   &  $<12.87$  &  $<12.72$  &  $<12.90$  &  $<12.90$  &  $<13.22$  \\
 & \oi{}  &  $<13.44$   &  $<14.10^g$  &  $<13.29$  &  $<13.44$  &  $<13.67$  &  $<14.30$  \\
 & \oi{}* $[\lambda1304]$ &  $<13.36$   &  $<13.32$  &  $<13.19$  &  $<13.36$  &  $<13.36$ &  $<13.67$  \\
  & \siii{}  &  $<12.41$   &  $12.38^{+0.11}_{-0.25}$  &  $\geq^f13.03$  &  $12.72^{+0.08}_{-0.17}$  &  $\geq13.20$  &  $\geq13.29$  \\
 & \siii{}*  &  $<12.03$   &  $12.33^{+0.21}_{-0.12}$  &  $12.61^{+0.13}_{-0.11}$  &  $12.75^{+0.09}_{-0.21}$  &  $12.99^{+0.08}_{-0.12}$  &  $13.10^{+0.08}_{-0.09}$  \\
 & \crii{}  &  $<12.62$   &  $<12.59$  &  $<12.47$  &  $<12.62$  &  $<12.61$  &  $<12.93$  \\
IV & \feii{} ($\,^6$D$_{9/2}$)  &  $13.17^{+0.09}_{-0.16}$   &  $13.75^{+0.08}_{-0.09}$  &  $12.66^{+0.13}_{-0.16}$  &  $12.86^{+0.14}_{-0.24}$  &  $13.07^{+0.10}_{-0.15}$  &  $13.92^{+0.06}_{-0.07}$  \\
 & \feii{}* ($\,^6$D$_{7/2}$)  &  $12.98^{+0.11}_{-0.32}$   &  $13.23^{+0.15}_{-0.12}$  &  $12.53^{+0.13}_{-0.12}$  &  $12.56$  &  $12.84^{+0.07}_{-0.05}$ &  $ 13.53\pm0.09$  \\
 & \feii{}** ($\,^6$D$_{5/2}$)   &  $12.67^{+0.13}_{-0.46}$   &  $12.94^{+0.15}_{-0.10}$  &  $<12.02$  &  $<12.57$  &  $<12.68$  &  $13.25^{+0.09}_{-0.10}$  \\
 & \feii{}*** ($\,^6$D$_{3/2}$)  &  $12.42^{+0.10}_{-0.45}$   &  $12.59^{+0.17}_{-0.08}$  &  $<12.23$  &  $<12.28$  &  $<12.56$  &  $12.93^{+0.10}_{-0.11}$  \\
 & \feii{}**** ($\,^6$D$_{1/2}$)   &  $12.29^{+0.18}_{-0.36}$   &  $12.36^{+0.24}_{-0.13}$  &  $12.23^{+0.15}_{-0.47}$  &  $<12.29$  &  $12.56^{+0.08}_{-0.16}$  &  $12.82^{+0.12}_{-0.13}$  \\
 & \feii{}***** ($\,^4$F$_{9/2}$) $[\lambda2348]$ &  $<12.97$   &  $<12.93$  &  $<12.81$  &  $<12.97$  &  $<12.97$  &  $<13.31$  \\
 & \feiii{} ($\,^5$D$_{4}$)  &  $13.53^{+0.26}_{-0.31}$   &  $14.02^{+0.13}_{-0.14}$  &  $13.78\pm0.17$  &  $13.87^{+0.17}_{-0.24}$  &  $14.13^{+0.13}_{-0.14}$  &  $14.44\pm0.09$  \\
 & \feiii{}* ($\,^5$D$_{3}$) $[\lambda1124]$ &  $<13.69$   &  $<13.65$  &  $<13.55$  &  $<13.69$  &  $<13.69$  &  $<13.95$  \\
 & \niii{} $[\lambda1370]$ &  $<13.23$   &  $<13.21$  &  $<13.06$  &  $<13.24$  &  $<13.24$  &  $<13.55$  \\
\rule[-0.2cm]{0mm}{0.4cm}
 & \znii{} $[\lambda2026]$ &  $<11.86$   &  $<11.83$  &  $<11.69$  &  $<11.86$  &  $<11.86$  &  $<12.19$  \\
\hline\hline
\end{tabular}
\vspace{-2mm}
 \tablefoot{$^a$ We adopt $z=2.42743$ as our zero-velocity reference point. $^b$ Purely turbulent broadening. $^c$ Logarithm of the ion column density and $1\sigma$ error estimate or $3\sigma$ upper limit. The detection criteria for each component are: \textit{i)} detection in at least both best-fit and upper continua; \textit{ii)} the column density is larger than the $3\sigma$ detection limit; \textit{iii)} its total error $err_{\textrm{tot}}= \sqrt{err_{\textrm{\begin{tiny}formal\end{tiny}}}^2 + err_{\textrm{\begin{tiny}normalization\end{tiny}}}^2 }<{\textrm{log}}\,2$. $^{d}$ Square brackets indicate the transitions used to measure the 3$\sigma$ upper limits. $^{e}$ The number of *'s indicates the excited energy level above the ground-state, either fine-structures of the ground-term or higher excited levels. $^f$ Conservative limits to account for any possible underestimate of \siii{} and \cii{} in component ``c'' in case of a low $b$-value. $^{g}$ Upper limit derived from deblending, see Sect.~\ref{sec oi}.}
\label{tab cd}
\end{table*}
necessary to cover the blended components, i.e., two FWHMs correspond to 27 \kms{} and 110 \kms{} for ``c+d'' and the total profile, respectively, corresponding to $b=16$ \kms{} and 66 \kms{}.

We applied the model fit derived above to the spectra collected at all the other epochs, allowing only the column density to change, in order to investigate any spectral variability. As a consistency check, an independent fit of epoch-II data confirms the positions and the widths of the components. 

Table~\ref{tab cd} lists the column densities derived from this four-component fit, for the four epochs. The summed components ``c+d'' and the sum of all the components are listed in the last two columns, for each level. We conservatively considered lower limits to the \cii{} content in component ``c'' because of the risk of the \cii{} $\lambda$1334 line being saturated with a $b$-value possibly lower than 7 \kms{} for this component. However, the non-detection of \cii{}* $\lambda$1335 - commonly detected in QSO damped \lya{} systems \citep[QSO-DLAs, log\,$N$(\hi{}) $>20.3$,][]{Wolfe86} - also indicates a low carbon content. In the same way, we conservatively considered lower limits to the \siii{} column densities of component ``c'' at epochs II and IV, when the stronger \siii{} $\lambda$1260 line was included. However, the \siii{} contents estimated from epoch-I and epoch-III data do not differ much from those of epochs II and IV, suggesting that any \siii{} underestimate must be small.

When summing upper limits, we used the non-significant estimates from the individual components when available\footnote{We only consider those non-significant individual-component column-density estimates that have an error smaller than $\sigma_{\log N}<1$, to avoid including measurements that are totally unconstrained.}, and otherwise expectation values of half the upper limits. This takes into account that individual components could be non-significant when considered singularly but that their sum may be $3\sigma$ significant. If none of the individual components to be combined have a non-significant column-density estimate, then the $3\sigma$ detection limit to either ``c+d'' or the total line profile is used. As noticed before, some transitions are covered at only two epochs.

 In addition, we summed the column densities of all the levels (i.e., ground-state and excited levels) belonging to the same ion (\oi{}, \siii{}, \feii{}, and \feiii{}) and list them in Table~\ref{tab cd_tot }. We again present the results for the individual and combined components.

\begin{center}
\begin{table}
\caption{Total ionic column densities including excited states.}
\begin{center}
\begin{tabular}{ @{}c@{\hspace{1.5mm}}|@{\hspace{1.5mm}}l@{\hspace{1.5mm}}|@{\hspace{1.5mm}}r >{\columncolor[gray]{ 0.9 }}  r@{\hspace{3mm}}  r@{\hspace{1.5mm}}|@{\hspace{1.5mm}}r@{} }
\hline \hline
\rule[-0.2cm]{0mm}{0.6cm}
Ep. & Ion & \multicolumn{4}{c}{log\,$N\pm\,\sigma_{\textrm{log}N}^a$}  \\
\rule[-0.2cm]{0mm}{0.6cm}
      &            &   comp. a   &  comp. b    & comps. c+d   &  \multicolumn{1}{c}{total} \\
\hline

\rule[-0.2cm]{0mm}{0.6cm}
I &  \feiii{} & $<14.46$ &  $14.08^{+0.22}_{-0.18}$ & $14.26^{+0.22}_{-0.18}$ & $14.58^{+0.17}_{-0.17}$ \\
\hline
\rule[0.0cm]{0mm}{0.4cm}
 &  \cii{} &  $<13.44$ &  $13.56^{+0.12}_{-0.15}$ & $\geq14.17$ & $\geq14.26$\\
II &  \oi{} &  $<14.02$ &  $<14.47^b$ & $<14.16$ & $<14.70$ \\
 &  \siii{} &  $12.74^{+0.16}_{-0.16}$ &  $12.91^{+0.10}_{-0.13}$ & $\geq13.44$ & $\geq13.62$ \\
\rule[-0.2cm]{0mm}{0.4cm}
 &  \feii{} & $13.72^{+0.08}_{-0.07}$ &  $14.34^{+0.03}_{-0.03}$ & $14.12^{+0.04}_{-0.04}$ & $14.60^{+0.03}_{-0.03}$ \\

\hline
\rule[-0.2cm]{0mm}{0.6cm}
III&  \feiii{} & $<14.04$ &  $14.13^{+0.13}_{-0.14}$ & $14.40^{+0.13}_{-0.13}$ & $14.61^{+0.10}_{-0.10}$ \\
\hline
 
\rule[0.0cm]{0mm}{0.4cm}
&  \cii{} &  $<13.23$ &  $13.13^{+0.18}_{-0.34}$ & $\geq14.12$ & $\geq14.17$\\
IV&  \oi{} &  $<13.70$ &  $<14.17^b$ & $<13.84$ & $<14.39$ \\
 &  \siii{} &  $<12.56$ &  $12.66^{+0.12}_{-0.13}$ & $\geq13.41$ & $\geq13.51$ \\
 \rule[-0.2cm]{0mm}{0.4cm}
 &  \feii{} & $13.59^{+0.07}_{-0.10}$ &  $13.97^{+0.07}_{-0.06}$ & $13.50^{+0.08}_{-0.07}$ & $14.20^{+0.05}_{-0.05}$ \\

\hline\hline
\end{tabular}
\tablefoot{$^a$ Logarithm of the ion column density and corresponding $1\sigma$ error estimate. $^{(b)}$ Upper limits from deblending. The highlighted column stresses the results for component ``b'', which shows the most peculiar [\siii{}/\feii{}].}
\label{tab cd_tot } 
\end{center}
\end{table}
\end{center}

\subsubsection{\crii{}}
\label{sec cr}
\begin{figure}
   \centering
   \includegraphics[width=60mm]{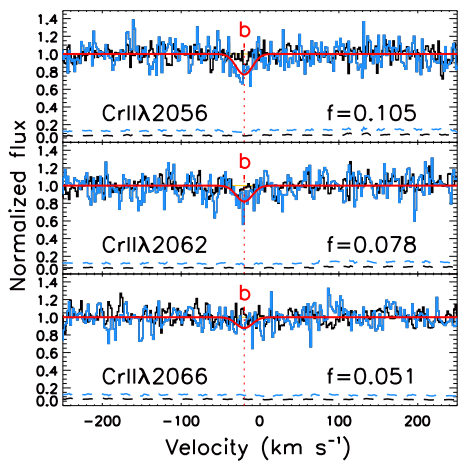}
      \caption{The \crii{} $\lambda\lambda\lambda$2056,2062,2066 transitions at epochs II (blue) and IV (black). The component ``b'' Voigt-profile fit at epoch II is shown by the solid red curve. The error spectra are displayed at the bottom of each panel with dashed lines.}
         \label{fig cr}
   \end{figure}

In addition to the low-ionization lines identified in the previous subsections, we identified \crii{} $\lambda\lambda$2056,2062 absorption features at $-20$ \kms{}, corresponding to component ``b'' in the line profile modelled for the other metals. We fitted the Voigt profile of component ``b'' (fixed $z$ and $b$-value) to these two transitions, together with the weaker \crii{} $\lambda$2066, deriving the column densities listed in Table~\ref{tab cd}. The other three components were not detected in the spectra and therefore we calculated the upper limit to the column densities as described above for the other transitions. We treated the Cr transitions separately from the other elements in order not to bias our results, as they are weak and only detected in component ``b''. Remarkably, the \crii{} column density for this component varied with time, at a 2.2 $\sigma$ level, as calculated from the \crii{} column density at epoch II and the expectation value on the upper limit at epoch IV. However, this is a conservative approach, as the errors in the column densities include the errors on the normalization. Using only the formal errors from the Voigt-profile fit, we derived a 4.6 $\sigma$ decrease in the \crii{} column density from epoch II to epoch IV (log\,$N_{\rm ep II}$(\crii{}) $ = 13.07 \pm 0.06$ and log\,$N_{\rm ep IV}$(\crii{}) $ = 12.37 \pm 0.14$). This is the first time that \crii{} absorption has been observed to vary in a GRB host galaxy. The spectral variability of \crii{} $\lambda\lambda\lambda$2056,2062,2066 is shown in Fig.~\ref{fig cr}.

\subsubsection{\hi{} and \oi{}}
\label{sec oi}

\begin{figure}
	\includegraphics*[width=89mm,angle=0]{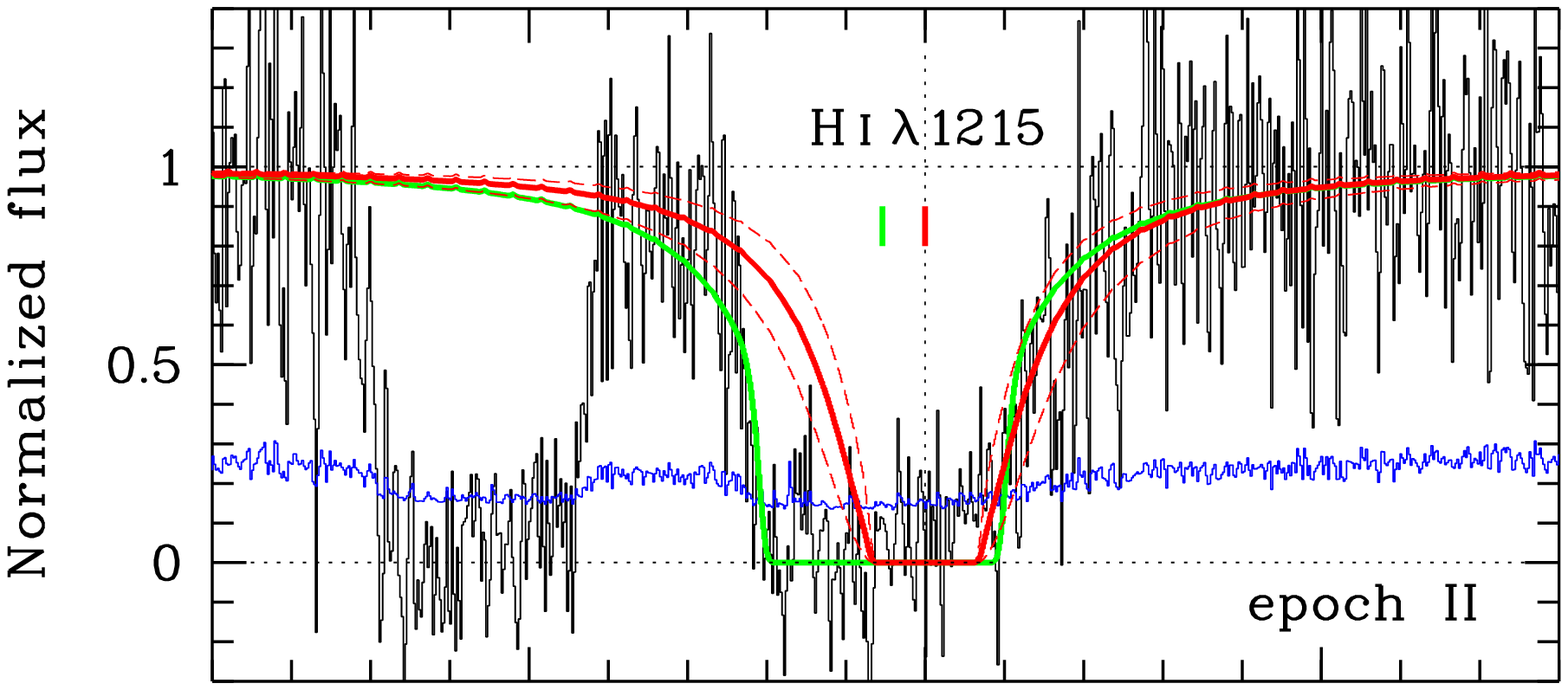}
 	\includegraphics*[width=89mm,angle=0]{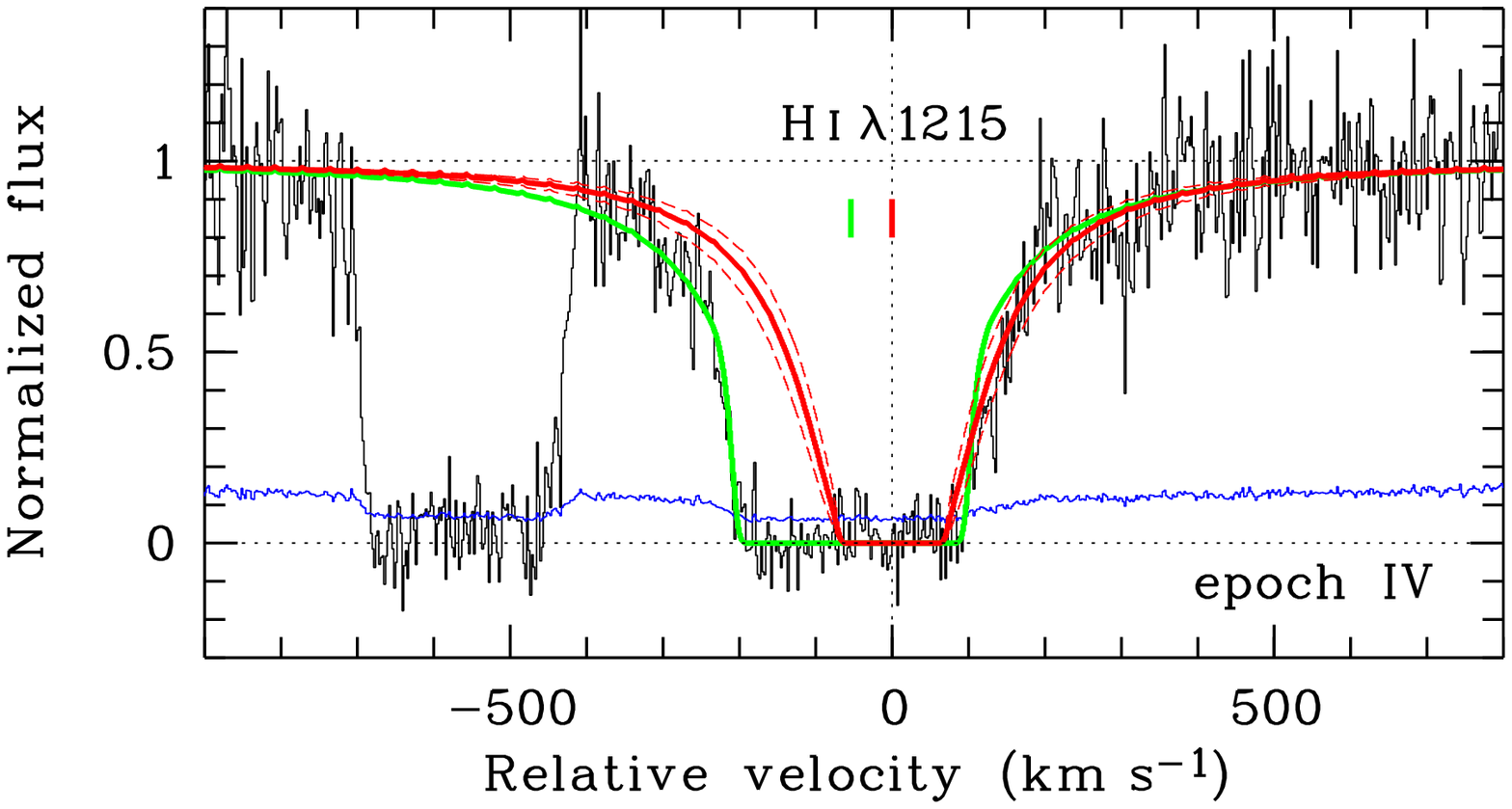}
      \caption{A neutral-hydrogen column-density fit to the damped \lya{} line at the host-galaxy redshift, shown with the red solid curves, provides log\,$N$(\hi{}) $ = 18.7 \pm 0.2$ at epoch II (top panel) and log\,$N$(\hi{}) $ = 18.7 \pm 0.1$ at epoch IV (bottom panel). The redshift was fixed to the mean redshift of the low-ionization metal components ``b'' and ``c'' ($z=2.427315$, taken here as the zero-point of the velocity scale), and the velocity broadening was fixed to $b=20$ \kms{}, corresponding to the difference in velocities of these two components or alternatively the quadratic sum of their $b$-values. The 1$\sigma$ errors are shown with dashed lines. To show that our measurement is robust, we also performed a fit - shown by the green curve - in which we allowed both the redshift and velocity broadening to vary as free parameters, finding $z=2.42670$, $b=46$ \kms{}, and log\,$N$(\hi{}) $ = 18.8$ for both epochs. The small vertical lines mark the central redshifts of the two independent fits.
}
         \label{fig Lya}
   \end{figure}

We derived the neutral-hydrogen column-density log\,$N$(\hi{}) $ = 18.7 \pm 0.2$ and log\,$N$(\hi{}) $ = 18.7 \pm 0.1$ at epochs II and IV, respectively, from a Voigt-profile fit of the \lya{} absorption line (Fig.~\ref{fig Lya}), using the MIDAS FITLYMAN package \citep{Fontana95}. These values were derived from the Voigt-profile fit of the red wing of \lya{}, fixing the central redshift and the $b$-value to the one found for the metal absorption system (mean of components ``b'' and ``c''). The extra absorption in the blue wing of the line could be related to either the GRB host-galaxy halo or galaxy winds observed in \civ{} \citep[see][]{Fox08}. As a sanity check, we also independently fitted the whole \lya{} profile with the redshift and $b$-value as free parameters. In this case, a larger $b$ was constrained by the extended core of the line, while $N$(\hi{}) is constrained by the wings of the line profile, providing log\,$N$(\hi{}) $ = 18.8$. Thus, there cannot be significantly more $N$(\hi{}) in this system than what is determined from the one-component fit at the redshift of the metal lines. This low column density classifies the absorber as a Lyman-limit system \citep[LLS,][]{Sargent89,Rauch98,Peroux03} with $17.3<$ log\,$N$(\hi{}) $<19$. Finally, we note that the absorption feature blue-shifted by roughly 600 \kms{} from \lya{} is probably due to \hi{} absorption along the line-of-sight and possibly related to the host galaxy. However, this component is unrelated to the metal absorption system that we focus on in the analysis.

The \oi{} $\lambda$1302 absorption associated with the GRB host galaxy is blended with the \alii{} $\lambda$1670 line from the $z=1.6711$ intervening system, listed in Table~\ref{tab inter}. The presence of \aliii{} absorption from the intervening system allows the \alii{} to be deblended from the GRB host-galaxy \oi{}. As the detected \aliii{} $\lambda\lambda$1854,1862 doublet is strong, and \feii{} and \siii{} are also detected at this redshift, a significant contribution from \alii{} is indeed expected. A total ionization of \alii{} can be excluded, its potential being higher than \feii{} and \siii{}.

At the $z=2.42743$ location of \oi{} $\lambda$1302, we observed a very weak absorption feature that may be associated with component ``c'' modelled as described above. However, the detection is formally not significant, given the noise level. No significant absorption was detected for components ``a'' and ``d'' either. Thus, we considered $3\sigma$ detection limits to the \oi{} column density for components ``a'', ``c'', and ``d'', listed in Table~\ref{tab cd}. To investigate how much \oi{} from the host-galaxy component ``b'' can be hidden in the \alii{} absorption associated with the intervening system, we first fitted the \alii{} and \aliii{} line profiles together with a Voigt profile, in the upper continuum level, for epoch IV, and we identified two components ``$\alpha$'' and ``$\beta$''. We then included the host-galaxy \oi{} contribution in the fit, resulting in a best-fitting $\chi_\nu^2= 1.12$ for log\,$N$(\oi{}, ``b'')$ = 13.50$ and the aluminium column densities listed in Table~\ref{tab alo} in the Appendix. The line profiles for the best fit are presented in Fig.~\ref{fig alo} in the Appendix. For a conservative estimate of the \oi{} column-density limit, we continued to increase the amount of oxygen in component ``b'', by allowing the \alii{} decrease until the reduced $\chi_\nu^2$ deviated more than $3\sigma$ from the best fit, resulting in the upper limit log\,$N$(\oi{}, ``b'')$ < 14.10$. We repeated this procedure for epoch II.

\subsubsection{\nv{}}
\label{sec nv}
Although \citet{Fox08} comprehensively discussed the high-ionization species along several GRB lines-of-sight, here we included \nv{} because we expected it to vary, given the \feii{} and \feiii{} variability observed here and the evidence, in some cases, that these transitions arise in the GRB vicinity \citep{Prochaska08b}. We analysed the \nv{} associated with the GRB host galaxy independently of the low-ionization species, since its high-ionization state associates it with a different gas phase with different kinematics, than the low-ionization species analysed in this paper.

The \nv{} $\lambda\lambda$1238,1242 doublet shows a clear two-component profile (see Fig.~\ref{fig: profiles var}), which we modelled using two Voigt profiles. We again used the highest S/N spectrum taken at epoch IV to derive a model-fit profile to apply to the spectra at other epochs. As a consistency check, we verified that a totally independent fit between epochs (i.e., leaving all the parameters free to vary at each epoch) did not change our results significantly. The \nv{} column densities at different epochs, summarized in Table~\ref{tab nv}, are consistent to within $1.1\sigma$, confirming the non-variation.

\begin{table}
\centering
\caption{\nv{} column densities and variability}
\begin{tabular}{ r | r r r }
\hline \hline
\rule[-0.2cm]{0mm}{0.8cm}
& \multicolumn{3}{c}{log\,$N$(\nv{})$\pm\,\sigma_{\textrm{log}N}^a$} \\
       &      blue comp.   &   red comp. &  total \\
\hline
\rule[-0.0cm]{0mm}{0.4cm}

Epoch  II   & 13.87$\pm$0.05 &     13.70$\pm$0.06 &  14.10$\pm$0.04 \\
\rule[-0.2cm]{0mm}{0.4cm}
Epoch  IV  & 13.81$\pm$0.02 &     13.68$\pm$0.03 &  14.05$\pm$0.02 \\
\hline
\rule[-0.2cm]{0mm}{0.6cm}
$\sigma_{II, IV}^b$  &  $1.1$ & $0.3$ & $1.1$ \\

\hline
\rule[-0.2cm]{0mm}{0.6cm}
$v^c$ (\kms{}) & $-53\pm1$ & $1\pm1$ & \\
\rule[-0.2cm]{0mm}{0.4cm}
$b$ (\kms{}) & $23\pm2$  & $18\pm2$& \\
\hline\hline
\end{tabular}
\tablefoot{$^a$ Logarithm of the \nv{} column density and corresponding $1\sigma$ error estimate, as derived from a Voigt-profile fit of the $\lambda\lambda$1238,1242 transitions ($\chi^2_\nu=1.126$ averaged between the two epochs). $^b$ Significance of the epoch II -- epoch IV variation. $^c$~Zero velocity reference at $z=2.42743$.}
\label{tab nv}
\end{table}

\subsection{Evidence for time variability}
\label{sec evidence var}

The \feii{} ground-state and fine-structure populations strongly decrease with time during the afterglow observations, as can be seen in Fig.~\ref{fig: profiles var} and Table~\ref{tab cd}. On the other hand, we observed a weak increase in the \feiii{} $^7$S$_3$ excited population, that is more pronounced in component ``b'', and in general more significant when we consider only the formal error in the Voigt-profile fit (i.e., not including the error in the normalization). While the \crii{} ground-state population shows hints of evolution with time, as presented in Sect.~\ref{sec cr}, we did not find any evidence of the variability of \cii{}, \siii{}, or \nv{} (see Sect.~\ref{sec nv}). In Table~\ref{tab var_sigma}, we report the significance level of variability in $\sigma$ units, for the total line profile and the individual component ``b'', including (and excluding) the error in the normalization.

\begin{center}

\begin{table}
\centering
\caption{Significance level of variability in $\sigma$ units}
\begin{tabular}{ l | c c }
\hline \hline
\rule[-0.2cm]{0mm}{0.8cm}
Population & $\sigma_{\rm var, b}$ & $\sigma_{\rm var, total}$ \\
\hline
\cii{}    &         1.8 (2.8)     &       0.6  \\
\siii{}    &         2.1 (2.5)     &        1.1\\
\siii{}*    &        0.2 (0.6)     &       0.2\\
\crii{}    &         2.2 (4.6)     &        2.2\\  
\feii{}    &         3.9 (3.9)     &        6.3\\
\feii{}*    &         2.5 (3.5)     &        3.9\\
\feii{}**    &         3.2 (5.3)     &        4.3\\
\feii{}***    &         2.3 (5.0)     &        2.8\\
\feii{}****    &         1.1 (2.7)     &        2.1\\
\feiii{}    &         1.7 (1.7)     &       0.6\\
\feiii{}$^{17}$*    &         3.2 (5.4)     &        2.5\\

\hline\hline
\end{tabular}
\tablefoot{$\sigma_{\rm var}$ between epoch II and epoch IV (I and III for \feiii{}$^{17}$*), calculated for component ``b'' and for the total line profile, considering the conservative larger errors that include the normalization error (the value in parenthesis considers only the formal errors from the Voigt-profile fit).}
\label{tab var_sigma}
\end{table}
\end{center}

\section{CLOUDY pre-burst photo-ionization modelling}
\label{sec cloudy}

To investigate the physical conditions and metallicity of the host-galaxy ISM prior to the burst, we ran a series of CLOUDY photo-ionization simulations \citep{Ferland98} designed to reproduce the observed ionic column densities. These time-independent (equilibrium) simulations do not apply once the GRB radiation has altered the ionization and excitation balance of the gas, but can be applied to the pre-burst column densities. These column densities were estimated from a photo-excitation/ionization modelling of the column density variability of this GRB afterglow performed in \citet[][paper II]{Vreeswijk12}, to produce the following results log\,$N$(\hi{}) $=18.7$, log\,$N$(\feii{}) $=14.8$, log\,$N$(\feiii{}) $=14.5$, log\,$N$(\siii{}) $=13.7$, and log\,$N$(\oi{}) $<14.6$, which we then used as input to the CLOUDY simulations.

\begin{figure*}[!ht]
\centering
\includegraphics[width=9.2cm]{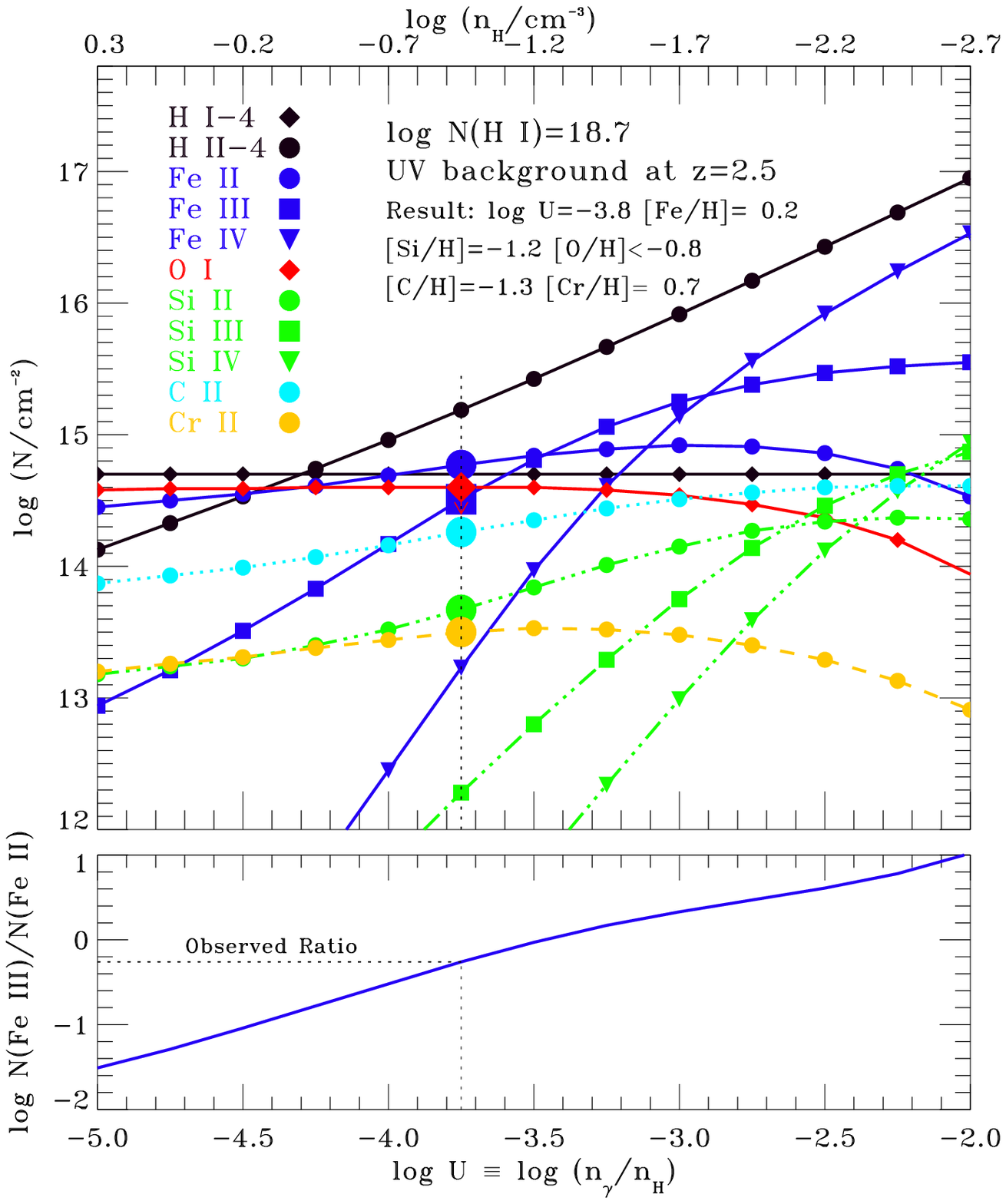}%
\includegraphics[width=9.2cm]{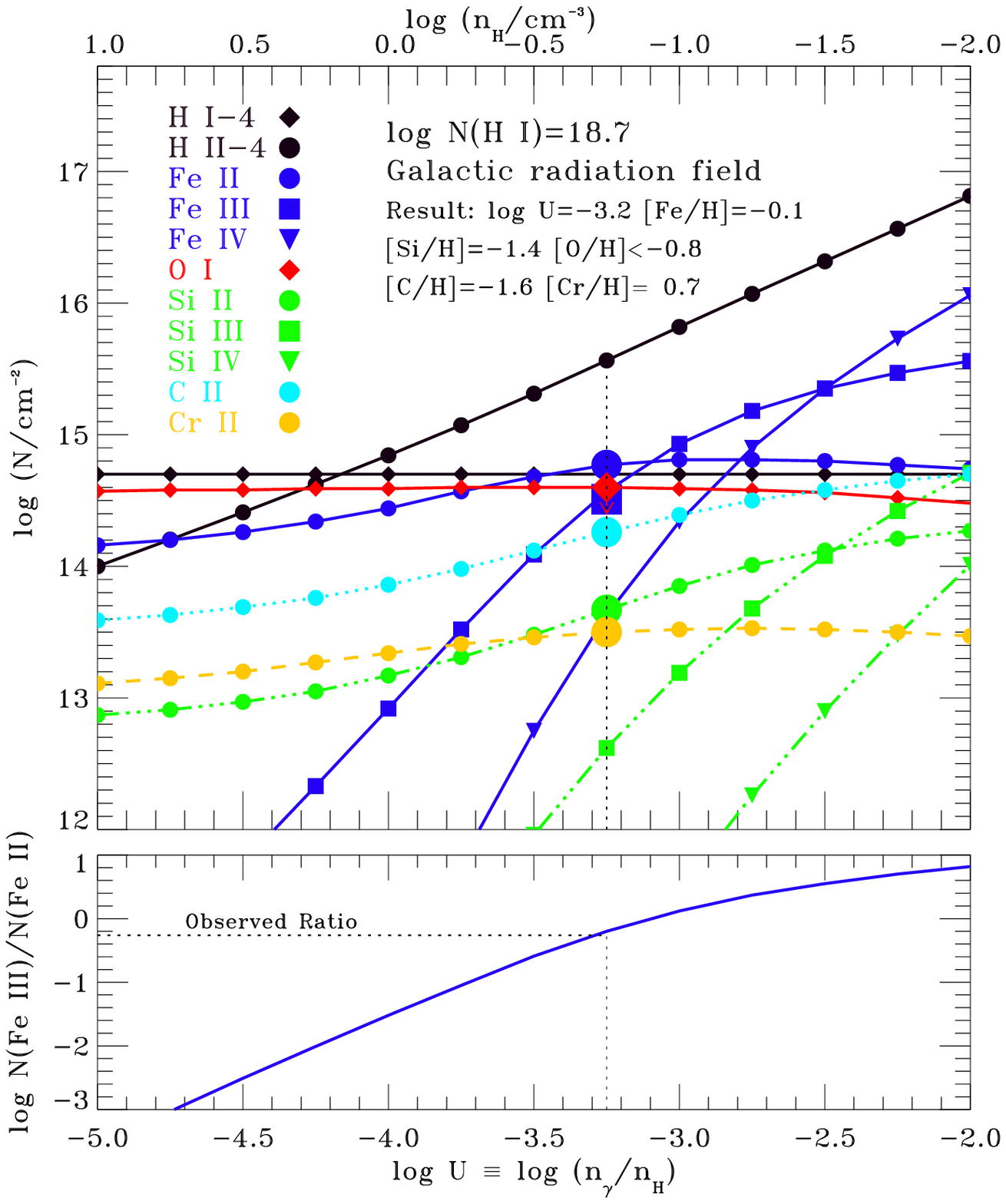}%
\caption{\textbf{Left:} CLOUDY photo-ionization model used to derive the pre-burst physical conditions and metallicity in the GRB 080310 host-galaxy ISM, assuming the extragalactic UV background radiation field at the GRB host-galaxy redshift. In the main panel, column density is plotted against ionization parameter $U$ for a range of metal lines. The observations are shown as large data points, colour-coded by ion. The best-fit log $U$ (and hence the best-fit density) and [Fe/H] are derived by matching the observed \feiii{}/\feii{} ratio, as shown in the bottom panel. [C/H], [O/H], [Si/H], and [Cr/H] are determined by matching \cii{}, \oi{}, \siii{}, and \crii{} at the best-fit log $U$. \textbf{Right:} Same as in the left panel but assuming the MW radiation field.} 
\label{fig cloudy}
\end{figure*}

The CLOUDY models assume that the gas exists in a plane-parallel uniform-density slab exposed to the $z=2.5$ extragalactic background (EGB) radiation incorporated within CLOUDY \citep[based on][]{Haardt96}, which has an ionizing photon density log\,($n_\gamma$/cm$^{-3})=-4.7$. We ran a grid of simulations for different values of the ionization parameter $U$, displayed in Fig.~\ref{fig cloudy} (left panel), where $U\!\equiv\!n_\gamma/n_{\mathrm H}$, the ratio of ionizing photon density to gas density. We then determined (a) the best-fit value of log $U$ by matching the observed \feiii{}/\feii{} ratio, which is a monotonically increasing function of log $U$ (see Fig.~\ref{fig cloudy}, bottom panel); (b) the best-fit value of [Fe/H] $=+0.2$ by matching the \feii{} and \feiii{} column densities at the best-fit log $U$; (c) the values of [C/H], [Si/H], and [Cr/H] and the upper limit to [O/H] by matching \cii{}, \siii{}, \crii{}, and \oi{} at the best-fit log U. The results are: $\log U = -3.8$, [C/H] $=-1.3$, [O/H] $<-0.8$, [Si/H] $=-1.2$, and [Cr/H] $=+0.7$. One clear result of this process, which we discuss in more detail below, is that the solar relative abundances fail to reproduce the column densities. We were able to estimate the error in the relative abundances to be $\sim0.2$, given the error in the pre-burst [\feiii{}/\feii{}] ratio of $\sim0.1$ dex (see paper II) and a similar uncertainty arising from the CLOUDY modelling.

To investigate the effect of a different radiation field on the results of our CLOUDY model, we ran the same grid of simulations as described above but assuming the Milky Way (MW) radiation field described in \citet{Fox05}. This should account for some star formation contributing to the UV radiation of the host galaxy. Although GRB host galaxies are generally highly star-forming (with the possible exception of \grb{}, as discussed in Sect.~\ref{discussion}), their mass is typically rather low \citep[on average $M=10^{9.3}M_\odot$;][]{Savaglio09}. Thus, using the integrated MW radiation field could provide a rough idea of the effect of star formation in the host galaxy on the ambient medium. The results of the alternative MW-radiation field CLOUDY model are shown in Fig.~\ref{fig cloudy} (right panel). In particular, the [Fe/H] and [Cr/H] ratios are both solar and super-solar for the two models, varying only little with the assumed radiation field. In addition, the [Si/Fe], [C/Fe], and [Si/Cr] ratios are similar - peculiarly low - in the two cases, independently of the model assumptions.

\section{Discussion}
\label{discussion}

\subsection{Low $N$(\hi{}) environment}

Most absorbers at the GRB redshift are DLAs, typically showing a higher $N$(\hi{}) than QSO-DLAs \citep{Vreeswijk04,Jakobsson06,Fynbo09}. This is a likely consequence of GRBs occurring in star-forming regions within their hosts, while examining the spectra of QSOs provides information about more random lines-of-sight through intervening galaxies \citep{Prochaska07}. Only a handful of {\it Swift} GRB absorbers are LLSs ($17.3<$ log\,$N$(\hi{}) $<19.0$), namely GRB\,050908, GRB\,060124, GRB\,060607A, \grb{} -- discussed in this paper --, and GRB\,090426. The \grb{} observations indicate that the line-of-sight does not cross any dense extended region with log\,$N$(\hi{}) $>19$.

In our dataset, \oi{} is in principle the most reliable metallicity indicator, as it is a low-ionization species (in the same phase and location as the low-ionization gas) not strongly depleted onto dust. Moreover, charge exchange between \oi{} and \hi{} \citep{Field71}, caused by the similarity of their ionization potentials, assures a constant ratio of the two species, independently of ionization effects. However, this is only valid for a cloud in equilibrium, i.e., embedded in a constant radiation field, and perhaps not for log\,$N$(\hi{}) $<19$ \citep{Viegas95}. Hydrogen and oxygen are ionized by the GRB afterglow at different rates in the case of neutral-hydrogen column-densities as low as log\,$N$(\hi{}) $\sim18.7$ (paper II).

From the above, caution should be exerted when using \oi{} to estimate the metallicity of the GRB host galaxy in LLS or lower \hi{} column density systems, as this could underestimate the overall metallicity. Moreover, the peculiar abundances of the \grb{} absorber cannot be simplified into one metallicity indicator (see Sect.~\ref{sec abundances}). 

Owing to the heavy saturation of the strong \lya{} transition, even at this low column density, the transition cannot be split into separate components in the same way as the metal lines, preventing a component-by-component analysis.

\subsection{\feiii{} origin and evidence for ongoing photo-ionization}
The \feiii{} UV\,34 $\lambda\lambda\lambda$1895,1914,1926 triplet, arising from the \feiii{} 17$^{\textrm{th}}$ excited level\footnote{See \feiii{} energy level diagram in \citet{Johansson00}.} ($^7$S$_3$), has never been reported in a GRB afterglow until now. However, it has been previously observed in the spectra of extreme environments such as broad-absorption lines BAL quasars and $\eta$~Carinae \citep{Hartig86,Wampler95,Baldwin96,Laor97,Johansson00,Hall02} in emission or absorption. In particular, these lines are detected in absorption in a rare class of iron low-ionization BAL quasars \citep[FeLoBAL, ][]{Becker97,Hall03}. In these environments \feii{} and \feiii{} excited levels are most likely populated via collisions, in a rather dense medium \citep[particle density $10^{10}<$ $n$(\hi{})/cm$^{-3}<10^{11}$, column density $\log N$(\hi{})$\sim21.3$,][]{deKool02}. However, the low $N$(\hi{}) that we measured for the \grb{} gas disfavours a collisional scenario\footnote{Even if the entire column would be compressed into a single parsec, the particle density would still only be $\sim2$ cm$^{-3}$.}, suggesting that a radiative process must be responsible for the population of the $\,^7$S$_{3}$ level. Moreover, the simultaneous variability of \feiii{} UV\,34 and \feii{} transitions favours a radiative process populating the \feiii{} excited levels. 

In $\eta$ Carinae, the \feiii{} UV 34 triplet was observed in \textit{emission} produced by the ion decaying from the $^7$P$^\circ$ term, $^7$P$^{\circ}_3$ being populated from the ground state via \lya{} pumping \citep{Johansson00}. From this level, the ion has a high probability of decaying to the $^7$S$_3$ level, and so in the presence of an increased \lya{} flux, this level can be populated efficiently. However, we note that this \lya{}-pumping channel cannot play a relevant role in a GRB environment, because both the GRB and pre-burst star-forming region cannot produce a significant \lya{} enhancement (pre-burst \lya{} photons cannot efficiently penetrate the absorbing cloud, while afterglow photons have little time to affect the line-of-sight \lya{}).

In paper II, we present a novel explanation of the \feiii{} UV\,34 triplet along the \grb{} line-of-sight: in this environment, the excited \feiii{} level $^7$S$_3$ is populated directly through the ionization of \feii{}.

We note that the variability of \feiii{} is stronger for component ``b'', while all the other features show a stronger variability in the total line profile (see Table~\ref{tab var_sigma}). This could suggest a higher rate of ionization of \feii{} into \feiii{} for component ``b'', while the variability in the other components could be dominated by photo-excitation. Together with the more extreme values of [Si/Fe] and [C/Fe] in component ``b'', this could be another hint that this component is the closest to the GRB. A component-by-component GRB-cloud distance estimate with a photo-ionization and -excitation model could either confirm or refute this hypothesis, but the difficulty in distributing \hi{} between the individual components would render the results uncertain.

In general, the observations of the \grb{} afterglow strongly indicate that we are witnessing the afterglow radiation gradually ionizing the surrounding medium. The pieces of evidence for such a photo-ionization in action are \textit{i)} the strong variability of the \feii{} populations from all energy levels, in particular the ground-state\footnote{Although \siii{} and \feii{} have very similar ionization potentials, their photo-ionization cross-sections \citep{Verner93,Verner96} are such that \siii{} is less ionized, which is consistent with the \grb{} observations.}; \textit{ii)} the detection and variability of \feiii{} transitions arising from the $^7$S$_3$ excited level, which are produced by ionization of \feii{}; \textit{iii)} the variability of the \crii{} ground-state population. The occurrence of photo-ionization in \grb{} is favoured by the low \hi{} content of the absorber, which does not efficiently shield the gas from the ionizing photons. However, it is mainly the high [Fe/H] that allows the \feiii{} UV34 triplet to be detected (see paper II).

\subsection{\nv{} non-variability}

The line profiles of the \nv{} doublet and the \feii{} fine-structure lines differ remarkably. In particular, the \nv{} lines display two components around 0 \kms{} and $-50$ \kms{}, and clearly lacks the component at $-20$ \kms{}. We do not detect any time variability in the \nv{} profiles (see Table~\ref{tab nv}), which would have been expected had the lines been produced within about 10 pc of the burst \citep{Prochaska08b}. Thus, we exclude a circumburst origin for the majority of \nv{} absorption in \grb{}. For QSO DLAs and sub-DLAs, \cite{Fox09} showed that \nv{} and \civ{} column densities are correlated and that their averaged $b$-values are similar, suggesting that the two ions arise in the same phase and have similar large-scale kinematics. Our analysis of the \nv{} line profiles and their non-variability favour the foreground host-galaxy ISM scenario for the production of \nv{} in the \grb{} afterglow.

\subsection{Relative abundances: ionization,~supernova~yields~or~dust~destruction?} 
\label{sec abundances}

\begin{table*}
\centering
\caption{Abundances towards \grb{} compared to other absorbers}
\begin{tabular}{l | r  r  r | c c c l}
\hline \hline
 & & & & \\
Ratio & comp. b & comps. c+d & total & \multicolumn{1}{c}{GRB DLAs} & \multicolumn{1}{c}{QSO (sub-)DLAs} &\multicolumn{1}{c}{QSO LLSs} & References \\
 & & & & \\
\hline

 & & & & \\
 
 $[$C/H$]$   & - & -  & $-1.3\pm0.2$         &                           &                           & $[-1.9,\,+0.2]$ &  [1]\\
 $[$O/H$]$   & - & -  & $<-0.8$      &                           &                           &                            &      \\
 $[$Si/H$]$   & - & - & $-1.2\pm0.2$         & $[-2.6,\,+0.7]$  & $[-2.6,\,+0.0]$  & $[-0.5,\,+0.0]$   &   [1],[2],[3],[4]\\ 
 $[$Cr/H$]$  & - & - & $+0.7\pm0.2$         &                           & $[-2.3,\,-0.8]$   &                             &  [3]\\ 
 $[$Fe/H$]$  & - & - & $+0.2\pm0.2$         & $[-3.0,\,+0.2]$  &  $[-3.0,\,+0.0]$  & $[-0.6,\,-0.4]$    & [2],[3],[5],[6],[7]\\
 $[$Ni/H$]$  & - & - &  $^a<+0.95$&                            &   $[-2.3,\,-0.6]$    &                          & [3]\\
 $[$Zn/H$]$ & - & - &   $^a<+1.08$& $[-1.8,\,+0.2]$  &   $[-2.0,\,+0.0]$   &                          &  [2],[4]\\
& & & & \\

\hline
 & & & & \\
$[$C/Fe$]$    & $\leq-1.74$           & $^{a,b}\leq-0.91$ &  $-1.5\pm0.2$        &                     & $[-0.4,\,+2.0]$ &                     & [7]\\
$[$O/Fe$]$    & $<-1.09$               & $<-1.18$             &  $<-0.8$     &                     & $[-0.4,\,+1.2]$ &                     & [7]\\
$[$Si/Fe$]$   & $\leq-1.47$           & $^{a,b}\leq-0.72$ &$-1.4\pm0.2$          & $[-0.3,\,+1.2]$ & $[-0.1,\,+0.7]$ &                     & [2],[3],[7],[8]\\
$[$Cr/Fe$]$ & $^a+0.56\pm0.19$ & $^a<+0.79$         & $+0.5\pm0.2$        &                     & $[-0.1,\,+0.5]$ &                     & [3],[8]\\
$[$Ni/Fe$]$ & $^a< +0.45$           &  $^a<+0.70$         & $^a<+0.52$ &                     & $[-0.2,\,+0.3]$ &                     & [3]\\
$[$Zn/Fe$]$ & $^a< +0.56$           &  $^a<+0.81$        &  $^a<+0.65$& $[+0.0,\,+1.7]$ & $[+0.0,\,+1.0]$ &                     & [2],[7]\\
& & & & \\

\hline\hline
\end{tabular}
\tablefoot{Abundances with respect to either H or Fe estimated for the \grb{} absorber (see text for a description of the pre-burst ionization corrections). The ranges of abundances observed in GRB or QSO absorbers are listed on the right side of the table. All these abundances are visualized in Fig.~\ref{fig ratios}. $^a$ Ionic relative abundances, i.e., not including the ionization effects of either the GRB or the ambient radiation field. $^b$ Does not include the possible -- albeit limited -- underestimate of Si and C in component ``c'' at epochs II and IV.}
\tablebib{[1] QSO LLSs, \citet{Prochaska99b}; [2] GRB DLAs, \citet{Schady11}; [3] \citet{Prochaska01}; [4] \citet{Wolfe05}; [5] QSO sub-DLAs, \citet{Dessauges09}; [6] QSO LLSs, \citet{Prochaska99}; [7] QSO sub-DLAs, \citet{Peroux03}; [8] \citet{Ledoux02}.}
\label{tab ratios}
\end{table*}

Estimating relative abundances from ionic column densities requires knowledge of ionization effects. Here, we first discuss relative \textit{ionic} abundances (e.g., [\siii{}/\feii{}]; similar to the definition of the relative abundances of chemical elements given in Sect.~\ref{sec intro}) for ions that are normally dominant in \hi{} clouds. We then refer to total element abundances after discussing the effects of ionization and deriving ionization corrections from CLOUDY modelling.

The relative \textit{ionic} abundances in the absorber have particularly low [\siii{}/\feii{}] $=-1.47\pm0.14$ and [\cii{}/\feii{}] $=-1.74\pm0.17$ ratios in component ``b'' ([\siii{}/\feii{}] $\geq-1.02$ for the total line profile and [\siii{}/\feii{}] $\geq-0.72$ in components ``c+d'') at epoch II. Despite the conservative lower limit to [\siii{}/\feii{}], we do not expect the ratio to be significantly underestimated (less than 0.1 dex), given the constraints on \siii{} at epochs I and III, so we can consider [\siii{}/\feii{}] $\sim-1.0\,\mbox{,}\sim-0.7$ for the total line profile and components ``c+d'', respectively. The observed [\siii{}/\feii{}] is difficult to explain since \siii{} and \feii{} have similar ionization potentials, as well as similar cosmic abundances. The typical value of [\siii{}/\hi{}] and high [\feii{}/\hi{}] (with a very low [\cii{}/\feii{}]) indicate that the low [\siii{}/\feii{}] is produced by an iron overabundance in the absorber rather than a silicon deficiency, which appears to be particularly marked in the peculiar component ``b''. We can exclude any significant fraction of \siii{} being hidden in higher-energy non-observable excited states, given the large energy gap between the first odd terms and the high decay probability \citep{Martin94}. Similarly, we can exclude a significant contribution of higher-excited states of oxygen because even if the non-detection limit for \oi{}** $\lambda$1306 is the same as for \oi{}* $\lambda$1304 (same oscillator strength), \oi{}** $\lambda$1306 is typically significantly weaker than \oi{}* $\lambda$1304.

\subsubsection{GRB ionization effects}

The larger overall photo-ionization cross-section of \feii{} relative to \siii{} \citep{Verner93,Verner96} indicates that \feii{} is ionized more effectively than \siii{} by the GRB afterglow. Therefore, the overabundance of \feii{} must have been even more pronounced before the onset of the burst. We found a pre-burst [\siii{}/\feii{}]~$=-1.1$ for the total line profile, using a photo-ionization and -excitation modelling of the GRB radiation on the surrounding medium (paper II; see Sect.~\ref{sec cloudy}).

\subsubsection{Pre-burst ionization effects}

The absorber being a LLS, ionization effects due to the host-galaxy and extragalactic radiation fields may influence the ionic abundances. At the low-ionization end, neutral species such as \fei{} and \sii{} are unexpected since they are easily ionized. In LLSs, \sii{} and \fei{} are expected to be more than 1000 times less abundant than their singly ionized species \citep{Tumlinson11}. This is confirmed by our CLOUDY photo-ionization modelling ($\log N(\mbox{\fei{}})<11$ for the best-fit ionization parameter).

Regarding the intermediate-ionization species, while we can measure \feiii{} to be at most as abundant as \feii{}, \siiii{} cannot be constrained from the spectra because \siiii{} $\lambda$1206 is blended with a strong \civ{} doublet associated with an intervening system. However, that \feii{} is more easily ionized than \siii{} reduces the possible amount of \siiii{} that could contribute to the overall Si content. This is also confirmed by the higher \feiii{}/\feii{} than \siiii{}/\siii{} (and \ciii{}/\cii{}) expected from an equilibrium photo-ionization modelling of LLSs \citep{Tumlinson11}, given the \grb{} \hi{} column density. Thus, [(\siii{}+\siiii{})/(\feii{}+\feiii{})] should be low as well. Additional strong support for this is provided by the low \cii{} content in component ``b''. Indeed, \cii{} can trace both \siii{} and \siiii{} because \cii{} has a much higher ionization potential than \siii{}. Therefore, we can exclude ionization effects such as the presence of \siiii{} as being the cause of the peculiar values of [\siii{}/\feii{}] and [\cii{}/\feii{}]. The results of our CLOUDY modelling confirm the low abundance of \siiii{} compared to \feiii{} (see Fig.~\ref{fig cloudy}).

On the high-ionization side, \ovi{}, \nv{}, \civ{}, and \siiv{} have different component structures and appear to arise in a separate ionization phase \citep{Fox08}. In addition, including higher ionization species in the abundance calculations would also require including \hii{}, which cannot be measured from the spectrum. Our CLOUDY modelling indicates that the pre-burst contribution of \feiv{}, \siiv{} - and thus also the high-ionization species with higher ionization potentials such as \civ{} and \ovi{} - to the total column densities is negligible.
 
Finally, a photo-ionization model for a LLS with N(\hi{}) and redshift similar to \grb{} (i.e., photo-ionized by the same UV background) confirms that for solar relative abundances, [\siii{}/\feii{}] is expected to be $\geq 0$ for all ionization parameters \citep{Prochaska99}. Thus, the low [\siii{}/\feii{}] observed in \grb{} is due to peculiar abundances rather than ionization. These ionization arguments limit the total abundances of Si and C relative to Fe, to [Si/Fe] $\leq$ [\siii{}/\feii{}] and [C/Fe] $\leq$ [\cii{}/\feii{}]. We rely on these estimates for the relative abundances of individual components.

For the total line profile, we can estimate the relative abundances with our CLOUDY photo-ionization model (Sect.~\ref{sec cloudy}), which is based on the pre-burst column densities derived in paper II. To assess how reliable these results are, we first note that these column densities are somewhat close to what we found at epochs I and II, when the burst had less time to ionize the surrounding medium, and they do not depend dramatically on the GRB photo-ionization model assumption (see paper II). Second, the relative abundances derived from the CLOUDY model for the adopted extragalactic background are similar to those found assuming the MW radiation field, strengthening our results. Thus, our estimates of the pre-burst ionization-corrected abundances are solid.

\begin{figure}
  	 \centering
 		\includegraphics[width=90mm,angle=0]{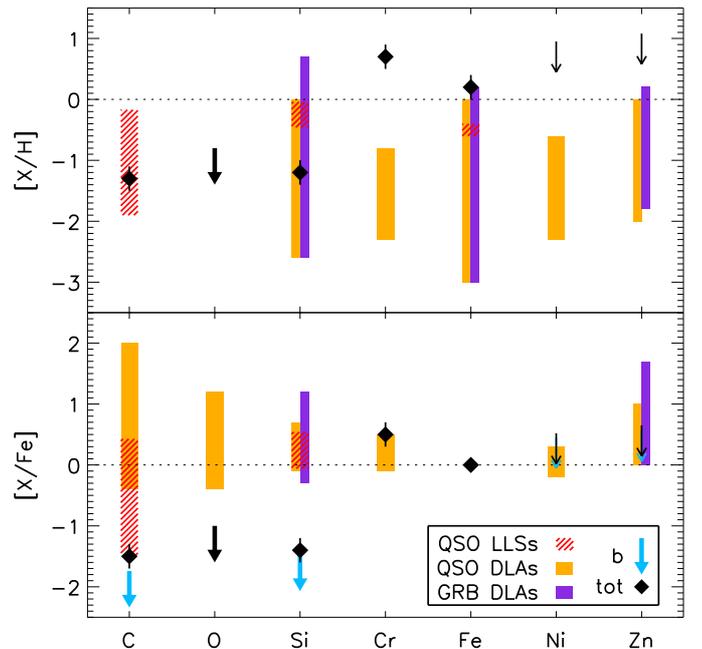}
      \caption{Abundances relative to H (top panel) and Fe (bottom panel) towards \grb{} (diamonds and arrows), compared to other absorbers (rectangular regions), taken from Table~\ref{tab ratios} (see text for a description of the pre-burst ionization corrections). The LLS [X/H] abundances are based on the observations of a few LLSs, and the corresponding [X/Fe] derived by combining the [X/H] ranges (the resulting [X/Fe] ranges are probably overestimated). Thin arrows are ionic abundances for \grb{}, i.e., uncorrected for ionization, observed at epoch II. The [C/Fe], [O/Fe], and [Si/Fe] estimated in \grb{} are much lower than previously observed in DLAs, sub-DLAs, and LLSs.}
         \label{fig ratios}
 \end{figure}

The relative abundances estimated in the \grb{} absorber - including the ionization correction discussed above - are listed in Table \ref{tab ratios}, along with literature values for large samples of both QSO-DLAs\footnote{We do not include the upper or lower limits to the QSO-DLA abundances listed by \citet{Prochaska01}. In particular, we exclude the outlier in the [Si/Fe] distribution, [Si/Fe] $>-0.617$ (J0255+00 absorber at $z=3.9$), because it has been estimated from a single Si line in the \lya{} forest (\siii{} $\lambda$1193) in a spectral region where the continuum level is poorly constrained.} and sub-DLAs, 3 LLSs, and 18 GRB absorbers. The same abundances are visualized in Fig. \ref{fig ratios}, suggesting that there is a heavy-metal enhancement in the absorbing gas, in particular Fe and Cr and possibly also Zn and Ni, given their high upper limits.

\subsubsection{Supernova yields}
\label{sec SN yields}

We have shown that the \grb{} line-of-sight exhibits peculiar abundances, and that these cannot be explained by ionization effects either owing to the afterglow or the extragalactic background radiation field. One possible explanation of the peculiar relative abundances, in particular the [Si/Fe] ratio, is that nearby supernovae (SNe) may have enriched the surrounding medium with heavy elements. Peculiar chemical abundances have occasionally been observed in QSO-DLAs and explained with supernova (SN) yields \citep[e.g.,][]{Agafonova11,Kobayashi11}. In particular, \citet{Agafonova11} suggested that the low [Si/Fe] $\sim-0.63$\footnote{This Si estimate should be taken with caution, because it is derived from \sii{} assuming that \sii{}/\siii{} $=$ \mgi{}/\mgii{}, which could lead to a significant Si underestimate, see photo-ionization models in, e.g., \citet{Tumlinson11}.} estimated in their $z=0.45$ absorber may be the remnant of the SN Ia explosion of a high-metallicity white dwarf \citep[using the SN Ia abundance patterns of][]{Stehle05,Mazzali08,Tanaka11}.

The metal yields from core-collapse SNe and hypernovae, including GRB-SNe, provide  $0<$ [Si/Fe] $<+2$, depending on the model \citep{Nomoto09}, failing to explain any sub-solar [Si/Fe] ratio. Models of Type Ia SNe predict [Si/Fe] ratios down to $\sim-0.2$ \citep{Maeda10} or as low as to $\sim-0.5$ \citep{Tanaka11} for Type Ia SN chemical enrichment, together with slightly super-solar [Cr/Fe] and [Ni/Fe] and low [Zn/Fe] ratios.

These predictions are partially consistent with the relative abundances observed in components ``c+d'' of the \grb{} absorber, suggesting that SNe Ia may have enriched the ISM with heavy metals along the line-of-sight. 

However, the extremely low [Si/Fe], [O/Fe], and [C/Fe] measured in component ``b'', as well as the pre-burst values inferred for the total line profile, cannot be reproduced only with SN Ia yields and require further explanation. \citet{Ohkubo06} found that a low [Si/Fe] $\sim-2.5$ could be expected for the chemical enrichment by the core-collapse of very massive Pop III stars (500--1000 $M_\odot$). However, these models also predict very high Ni and Zn contents ([Ni/Fe] $\sim+1.0$, [Zn/Fe] $\sim+1.6$) and a strong depletion of chromium ([Cr/Fe] $\sim-1.0$), all features that we do not observe. Exotic processes such as ONeMg core explosions or accretion-induced white-dwarf collapses could qualitatively explain these peculiar abundances, but they could only contribute marginally to the ISM abundances.

In general, the low [Si/Fe] observed in the \grb{} absorber also requires no $\alpha$-element contributions from core-collapse SNe, indicating that there has been a small amount of recent star formation in the host galaxy. Thus, a long GRB \citep[likely arising from a massive star progenitor, e.g.,][]{Woosley93,Hjorth11} indicates episodic massive-star formation in the host galaxy. 

The lack of recent star formation along the line-of-sight is perhaps another indication - after the low $N$(\hi{}) - that \grb{} is not located within the densest parts of its host galaxy, given that GRB hosts are typically actively star-forming \citep[e.g.,][]{Savaglio09}. The evidence of some GRB absorbers - those with almost featureless continua - that are not deeply embedded in the densest regions of their host galaxies was discussed in \citet{DeCia11b}. However, several, strong absorption lines are detected in the case of \grb{}. This could suggest that the LLS environment of \grb{} differs intrinsically from those of most GRB-DLAs, corresponding to possibly a neutral-gas poor and not highly star-forming host galaxy, and that the measurements are not simply due to a particular location of the burst within its host galaxy.

\subsubsection{Dust destruction}

One intriguing possibility for explaining the Fe and Cr enhancements in component ``b'' could be dust destruction, possibly induced by the GRB itself. From the Voigt-profile modelling, \feii{} and \crii{} show stronger absorption in component ``b'', suggesting that these ions are mostly co-spatial. Moreover, Fe and Cr are the most depleted elements in the dust grains of QSO- and GRB-DLAs \citep[e.g.,][]{Savaglio03}. The destruction of silicate grains can release Fe and recycle it into the ISM out to at least 3 pc from the GRB, increasing $N$(Fe) by one order of magnitude and not affecting the abundance of less depleted elements such as O \citep{Perna02}. In addition, Fe and Fe$_3$C are the first and sometimes the only grains to condensate at high temperature \citep[$T\sim1000$~K,][]{Lewis79}.

If dust destruction recycles the metals into the ISM, the relative abundances of the gas no longer suffer from dust depletion and represent the intrinsic abundances. This alone does not explain the peculiar Fe and Cr overabundances that we observe. However, this only applies when the gas and dust evolve together, which is not necessarily the case. In particular, radiation pressure can cause the dust grains to drift away from the inner parts of \hii{} regions \citep{Draine11}. Large dust grains can also be driven out to at least beyond 0.5 pc into the ISM by stellar-wind shocks \citep{VanMarle11}. Both of these processes can therefore change the local dust/gas ratio within star-forming regions. The iron overabundance in component ``b'' may be the result of the destruction of Fe-rich dust grains, such as iron silicates or solid Fe, while the low C abundance suggests that this destroyed dust cloud has a low carbonaceous content.

The GRB radiation can quickly destroy dust grains out to a distance of $\sim$100 pc from the burst \citep{Fruchter01}. 
Alternatively, a GRB-unrelated process such as shocks (e.g., in SN ejecta) could be responsible for the dust destruction \citep[e.g.,][]{Jones11,Jones96}. This has never been reported in QSO absorbers, but QSO lines-of-sight typically probe the outer regions of the intervening galaxies, where a low dust content and SN rate are expected. In conclusion, the destruction of drifted Fe-rich dust grains, possibly due to the burst or a nearby SN, could explain the iron overabundance in component ``b'', but further modelling is clearly needed to investigate the feasibility of this scenario.

In any case, a large amount of dust in the absorber can be excluded, because if a significant amount of iron would be depleted into dust then the intrinsic [Si/Fe] would have to be even lower.

\section{Conclusions}

Our time-series of high-resolution VLT/UVES spectra of the \grb{} afterglow has revealed the unique features of the ISM of the GRB host galaxy. We have reported the detection of several resonance absorption lines commonly observed in the ISM, as well as \feii{} fine-structure lines sometimes associated with GRB afterglows. In the case of \grb{}, both the \feii{} ground-state and the fine-structure lines vary with time. We decomposed the complex spectral-line profiles of several ions and modelled them altogether using a four-component Voigt profile, resulting in column-density determinations for each component. These components might be associated with different clouds along the line-of-sight within the GRB host galaxy. Interestingly, a low $\log N$(\hi{}) $=18.7\pm0.1$ is derived from the \lya{} absorption.

We also detected the \feiii{} UV\,34 $\lambda\lambda\lambda$1895,1914,1926 line triplet in absorption. These transitions arise from the $^7$S$_3$ energy level of \feiii{}, which is observed for the first time in a GRB afterglow spectrum. The \feii{} and \feiii{} time variability, of both the ground-state and excited levels, is clear evidence that we are witnessing the ISM being gradually photo-ionized by the GRB afterglow radiation.

Through the analysis of chemical abundances measured in the \grb{} absorber and CLOUDY photo-ionization modelling, we inferred pre-burst ionization-corrected ratios of [C/Fe] $=-1.5\pm0.2$, [O/Fe] $<-1.0$, and [Si/Fe] $=-1.4\pm0.2$ (where [Fe/H] $=+0.2\pm0.2$ and [Cr/H] $=+0.7\pm0.2$), while, typically, $0\lesssim$ [Si/Fe] $\lesssim+0.7$ is observed in QSO and/or GRB absorbers. Furthermore, we observed an even more extreme iron overabundance ([Si/Fe] $\leq-1.47$ and [C/Fe] $\leq-1.74$) in the gas associated with component ``b'' of the absorption profile. Such a low [Si/Fe] ratio has never been observed before in either QSO- or GRB-DLAs and cannot easily be explained by current models of SN Ia and Pop~III SN chemical yields. A potential explanation might be provided by the destruction of iron-rich dust grains, thereby recycling heavy elements into the gas phase. The dust could be destroyed by the GRB itself or GRB-unrelated processes such as SNe shock waves.

The high iron column-density measured in the gas phase generally suggests a low dust content in the absorber. The strong overabundance of iron compared to silicon and carbon also suggests that there has been negligible recent star formation along the line-of-sight. The occurrence of the GRB then indicates that there has been episodic massive-star formation in the GRB region.

\begin{acknowledgements}
We thank Peter Laursen and Keiichi Maeda for insightful discussions and Jason Prochaska and Sandra Savaglio for useful comments. We also thank an anonymous referee for a careful and constructive report. ADC acknowledges support from the ESO Director General Discretionary Fund 2009 and 2010, and the University of Iceland Research Fund. PJ acknowledges support by a Marie Curie European Re-integration Grant within the 7th European Community Framework Program and a Grant of Excellence from the Icelandic Research Fund. The Dark Cosmology Centre is funded by the Danish National Research Foundation. We acknowledge the careful assistance of the VLT observers, in particular Claudio Melo and Dominique Naef. 
\end{acknowledgements}

\bibliographystyle{aa} 

\bibliography{biblio}

\appendix

\section{}

\begin{figure*}
   \centering
  	\includegraphics{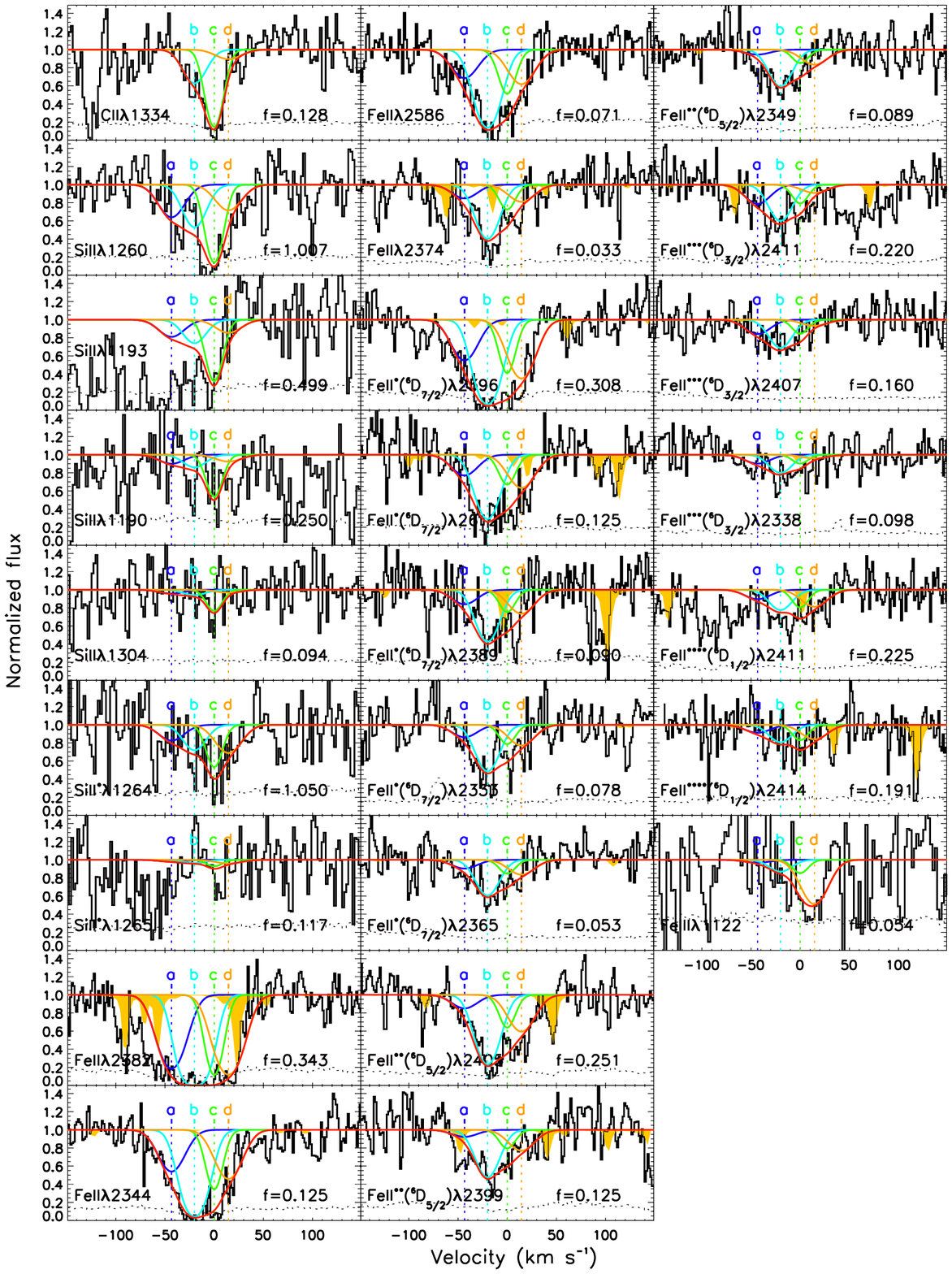}
     \caption{Voigt-profile fit of absorption lines observed at epoch II. The fit decomposition was derived from the highest S/N spectrum (obtained at epoch IV).}
         \label{fig: fit 2}
   \end{figure*}

\begin{figure*}
   \centering
  	\includegraphics{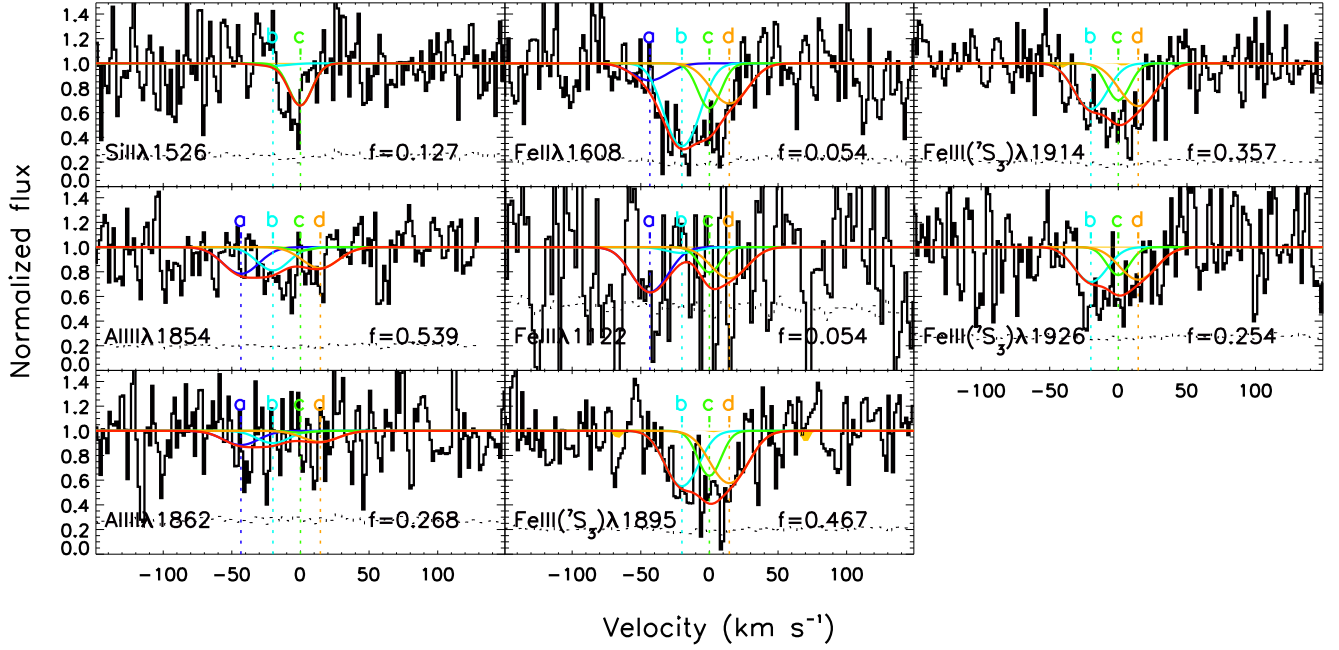}
     \caption{Voigt-profile fit of absorption lines observed at epoch I. The fit decomposition was derived from the best S/N spectrum (obtained at epoch IV). For the most poorly resolved lines, e.g., \feiii{} $\lambda$1122, we conservatively adopt upper limits, see Table~\ref{tab cd}.}
         \label{fig: fit 1}
   \end{figure*}

\begin{table*}
\centering
\caption{Intervening systems: identified absorption lines}
\begin{tabular}{@{}c | r | c | c  c c c r@{}}
\hline \hline
\rule[-0.3cm]{0mm}{0.8cm}
 $z$  &  $v_{\rm host}^a$& Transitions & \multicolumn{5}{c}{$W_{\rm rest}^b$ (\AA{})}\\
        & (\kms{}) & $\lambda$ (\AA{}) & \lya{} & \civ{} 1548 & \mgii{} 2796 & \siii{} 1526 & \feii{} 2382\\
\hline                     
\rule[-0.0cm]{0mm}{0.4cm}
  2.4200 & $650$& \lya{}, \civ{} 1548, 1550$^c$, & $1.04\pm0.01$  &$0.05\pm0.01$ &$<0.05$ &$<0.02$ & $0.05\pm0.01$\\
  & & \feii{} 2382 & & & & & \\ 
  
\hline  
\rule[-0.0cm]{0mm}{0.4cm}
  2.4113 & $1411$&\lya{}, \civ{} 1548, 1550$^c$ &$0.70\pm0.01$  &$0.13\pm0.02$ & $c$ & $<0.03$& $<0.02$\\
   
  \hline
\rule[-0.0cm]{0mm}{0.4cm}
  2.2786 &$13018$ &\lya{}, \cii{} 1334, \cii{}* 1335,  & $1.62\pm0.01$ ($\approx18.7^d$) &$1.03\pm0.02$ & $0.07\pm0.02$& $<0.02$& $<0.02$\\
  & &  \civ{} 1548,\,1550, & & & & & \\
  & &\mgii{} 2796, 2803, & & & & & \\
         & & \siii{} 1260$^c$, \siiii{} 1206,  & & & & & \\
  & &  \siiv{} 1393, 1402, & & & & & \\    
         &  &\alii{} 1670, \aliii{} 1854, 1862 & & & & & \\
  
  \hline       
\rule[-0.0cm]{0mm}{0.4cm}
  2.1702 &$22499$ & \lya{}, \civ{} 1548, 1550 & $0.61\pm0.01$ & $0.07\pm0.01$&$c$ & $<0.02$& $<0.02$\\
 
 \hline 
\rule[-0.0cm]{0mm}{0.4cm}
  2.0685 &$31395$ &\lya{}, \civ{} 1548, 1550 & $1.22\pm0.02$ ($\approx18.4^d$) & $0.21\pm0.01$& -- &$<0.02$ & $<0.02$\\
  
  \hline
\rule[-0.0cm]{0mm}{0.4cm}
  1.6711 & $66155$& \lya{}, \cii{} 1334, \cii{}* 1335,  & $<0.96$ & $c$&$0.42\pm0.01$ &$0.13\pm0.01$ & $0.04\pm0.01$\\ 
  & &  \civ{} 1548$^c$,\,1550,  & & & & & \\ 
           & &\mgii{} 2796, 2803, & & & & & \\
          &  &\alii{} 1670, \aliii{} 1854, 1862, & & & & &  \\
         & &\siii{} 1526, \siiv{} 1393, 1402, & & & & & \\
         & &  \feii{} 1608, 2344, 2382, 2600 & & & & & \\
 
 \hline       
\rule[-0.2cm]{0mm}{0.6cm}
  1.1788 & $109216$&\siii{} 1526, &  & $<0.18$& $0.05\pm0.01$& $0.24\pm0.04$& $<0.04$\\
         & &   \mgii{} 2796, 2803 & & & & & \\

\hline \hline
\end{tabular}
\tablefoot{$^{a}$ Blue-shifted velocity with respect to the $z=2.42743$ host-galaxy system. $^b$ Rest $W$ and $1\sigma$ error estimate, or $3\sigma$ upper limit for non-detections, measured in the highest S/N spectrum for each line. The assumed 2$\times$FWHM apertures correspond to a $b$-value of 20 km s$^{-1}$, which enabled us to cover each line (except for \civ{} 1548 at $z=2.41, 2.27,2.06,1.67$, where an aperture corresponding to $b=100$ km s$^{-1}$ was necessary and $b=50$ km s$^{-1}$ for \mgii{} 2796 at $z=1.67$ and \civ{} 1548 at $z=1.17$). $^{c}$ Blended with another line. $^d$ $\log N(\mbox{\hi{}})$ estimate from the $W$, assuming a damped regime ($N_{\mbox{\lya{}}}[\mbox{cm}^{-2}]\approx 1.87 \times 10^{18} W^2[\mbox{\AA{}}]$)}
\label{tab inter}
\end{table*}

\begin{table*}
\caption{Intervening system at $z=1.6711$: column densities of \alii{} and \aliii{}}
\centering
\begin{tabular}{ c | c c }
\hline \hline
\rule[-0.2cm]{0mm}{0.8cm}
Ion  & \multicolumn{2}{c}{log\,$N\pm\,\sigma_{\textrm{log}N}^a$}  \\
\rule[-0.2cm]{0mm}{0.4cm}
       &    $\alpha$ & $\beta$\\
\hline
\rule[-0.0cm]{0mm}{0.4cm}
\alii{} & $11.33\pm0.26$  & $12.26\pm0.07$\\
\rule[-0.2cm]{0mm}{0.4cm}
\aliii{}&  $12.28\pm0.19$ & $13.01\pm0.04$ \\
\hline
\rule[-0.0cm]{0mm}{0.4cm}
$z$ & $1.671110\pm0.000007$ & $1.670923\pm0.000029$\\
\rule[-0.2cm]{0mm}{0.4cm}
$b$  & $8\pm5$ \kms{}& $9\pm1$ \kms{}\\
\hline\hline
\end{tabular}
 \tablefoot{$^a$ Logarithm of the ion column density and corresponding $1\sigma$ error estimate, for components $\alpha$ and $\beta$.}
\label{tab alo}
\end{table*}

\begin{figure*}
   \centering
   \includegraphics[width=85mm]{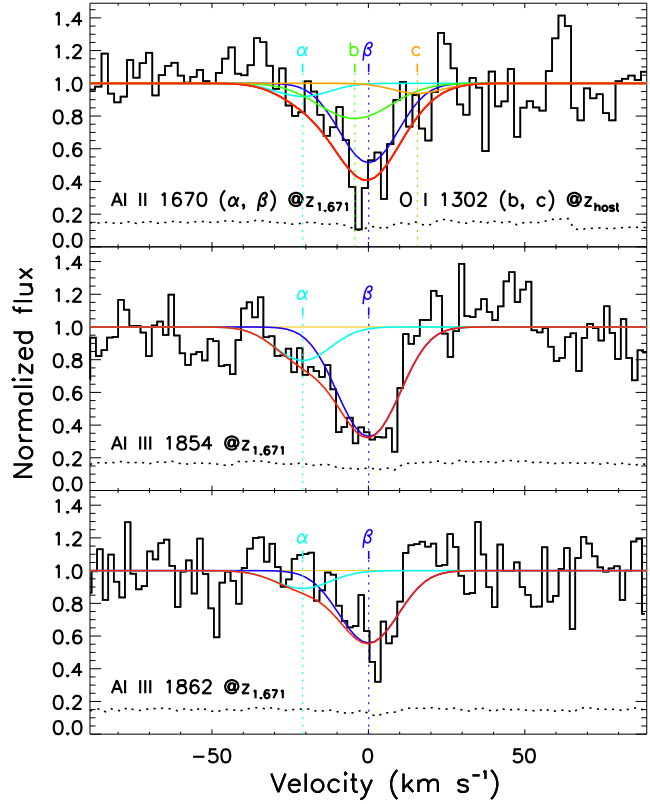}
      \caption{Voigt-profile fit of \alii{} and \aliii{} -- associated with an intervening system -- and the host-galaxy \oi{}, observed at epoch IV, assuming the best-fitting log\,$N$(\oi{},``b'')$=13.50$ value. The \alii{} $\lambda$1670 line at $z=1.6711$ is blended with \oi{} $\lambda$1302 at the GRB host-galaxy redshift. The \oi{} line estimate is discussed in Sect.~\ref{sec oi}.}
         \label{fig alo}
   \end{figure*}

\end{document}